\documentclass[prb,showpacs,preprintnumbers,amsmath,amssymb,
floatfix,superscriptaddress]{revtex4}
\usepackage {rotating}
\usepackage{graphicx}
\usepackage{dcolumn}
\usepackage{ulem,color}
\begin{document}

\preprint{Preprint 01} 

\title{
    Structure and dynamics of oxygen adsorbed on Ag(100) vicinal
    surfaces
}
\author{N. Bonini}
\affiliation{%
  SISSA -- Scuola Internazionale Superiore di Studi Avanzati and 
  INFM DEMOCRITOS National Simulation Center, Via Beirut 2-4,
  34014 Trieste, Italy
}%
\author{A. Kokalj}
\affiliation{%
  SISSA -- Scuola Internazionale Superiore di Studi Avanzati and 
  INFM DEMOCRITOS National Simulation Center, Via Beirut 2-4,
  34014 Trieste, Italy
}%
\affiliation{%
  Jo\v zef Stefan Institute, 1000 Ljubljana, Slovenia
}%
\author{A. Dal Corso}
\affiliation{%
  SISSA -- Scuola Internazionale Superiore di Studi Avanzati and 
  INFM DEMOCRITOS National Simulation Center, Via Beirut 2-4,
  34014 Trieste, Italy
}%
\author{S. de Gironcoli}
\affiliation{%
  SISSA -- Scuola Internazionale Superiore di Studi Avanzati and 
  INFM DEMOCRITOS National Simulation Center, Via Beirut 2-4,
  34014 Trieste, Italy
}%
\author{S. Baroni}
\affiliation{%
  SISSA -- Scuola Internazionale Superiore di Studi Avanzati and 
  INFM DEMOCRITOS National Simulation Center, Via Beirut 2-4,
  34014 Trieste, Italy
}%

\date{\today}

\begin{abstract}
  The structure and dynamics of atomic oxygen adsorbed on Ag(410) and
  Ag(210) surfaces has been investigated using density functional
  theory. Our results show that the adsorption configuration in which
  O adatoms decorate the upper side of the (110) steps forming
  O--Ag--O rows is particularly stable for both surfaces. On Ag(210),
  this arrangement is more stable than other configurations at all the
  investigated coverages. On Ag(410), adsorption on the terrace and at
  the step edge are almost degenerate, the former being slightly
  preferred at low coverage while the latter is stabilized by
  increasing the coverage. These findings are substantiated by a
  comparison between the vibrational modes, calculated within
  density-functional perturbation theory, and the HREEL spectrum which
  has been recently measured in these systems.
\end{abstract}

\maketitle

\section{Introduction} \label{sec:int}
The interaction of oxygen with silver surfaces has been intensively
studied because of the key role of silver in many important industrial
oxidation reactions such as, for example, partial oxidation of
methanol to formaldehyde or ethylene epoxidation.\cite{vanSanten_AC35}
These reactions occur very efficiently when Ag powders are employed as
a catalyst. Since the particles which compose these powders
are affected by many surface defects of different nature and with
different concentrations, their chemistry can be extremely different
from that of perfect low-Miller-index surfaces. Indeed the presence of
various inequivalent adsorption sites suggests that atomic and
molecular oxygen can adsorb forming species of different chemical
nature. The microscopic characterization of such adsorbed species is a
fundamental issue to understand the catalytic activity of Ag powders.

A large amount of experimental and theoretical results clearly
indicates that steps, kinks, and other surface imperfections play a
crucial role in surface chemistry. For example, the study of many
systems such as N and O adsorbed on Ru(0001),\cite{hamm00} O on
Pt(111),\cite{feib96} or C$_2$H$_4$ on Ag(100)\cite{koka02} has shown
that monoatomic steps can bind adsorbates more strongly than terrace
sites. Moreover it has been shown that this kind of defects can
completely determine the kinetics of dissociation in systems such as,
for example, N$_2$ and NO on Ru(0001),\cite{dahl99,zamb96} O$_2$ on
Pt(111),\cite{gamb01} N$_2$ on Fe(110) and Fe/Ru(0001).\cite{egeb01}

The role of steps in the dynamics of O$_2$/Ag interaction has been
recently investigated by Rocca and
coworkers\cite{savi01,savi02,vatt03} who have studied the Ag(410) and
Ag(210) surfaces. These surfaces are vicinal of Ag(100) and are
characterized by open (110) steps and three-atom- and one-atom-row
wide (100) terraces, respectively. Using a supersonic molecular beam
to dose O$_2$ onto Ag surfaces at selected angles of incidence and
different kinetic energies, and characterizing the adsorbed species by
vibrational spectroscopy (HREEL), these authors have investigated the
various oxygen species which form on these surfaces. It was concluded
that O$_2$ dissociation occurs mainly at the steps and that atomic
oxygen adsorbs at different surface sites. In particular, HREEL
spectra show two peaks in the frequency region of the O/Ag stretch
modes, at about 32 and 40 meV. These frequencies are quite similar to
those observed in O/Ag(100) (33 meV)\cite{baut99} and on the added-row
reconstructed O/Ag(110) surface (40 meV).\cite{vatt94} It was thus
proposed to assign the low energy peak to adatoms occupying (100)
terrace sites and the one at higher energy to atomic oxygen adsorbed
at the step edges forming O--Ag--O chains similar to the added rows
found in reconstructed O/Ag(110) surfaces.\cite{cour97,schi93} In
O/Ag(210) an additional peak around 56 meV was found, and it was
proposed that it is due to the vibration of an oxygen atom occupying a
subsurface site.

Motivated by these results we have performed {\it ab initio}
calculations aimed at understanding the role of the steps in the
adsorption of atomic oxygen on the Ag(410) and Ag(210) surfaces. In
the present work we limit our investigation to on-surface atomic
oxygen, subsurface and molecular adsorption will be addressed in
separate papers. Our results show that on these vicinal surfaces the
adsorption configuration in which the adatoms decorate the steps in a
(1$\times$1) geometry is particularly stable. The oxygen adsorption on
the two surfaces presents however significant differences related to
the availability of adsorption sites on terraces. Indeed, on Ag(410),
terrace and step edge sites have comparable chemisorption energies
and, in particular, the hollow sites on terraces are slightly more
favored than the step edge sites when the adatoms are far apart from
each other.  At higher coverage, instead, adatoms slightly prefer to
decorate the steps. On Ag(210), the absence of favorable terrace sites
makes the step decoration much more stable than any other adsorption
configurations.  The formation of these O--Ag--O chains at step edges
will be shown to significantly affect both the geometrical structure
and the electronic properties of the surface.  As we have already
mentioned, the occurrence of adsorption on both the terraces and steps
of vicinal Ag(100) surfaces was also proposed based on an analysis of
the HREEL spectra, recently measured for these
systems.\cite{savi01,savi02,vatt03} Our calculation of the vibrational
properties of these systems---performed within density-functional
perturbation theory---provides a further support to this analysis.

The paper is organized as follows. In Section \ref{sec:theo} we
describe the theoretical approach and computational details of our
work. Section \ref{sec:res} contains our results for the energetics,
the geometrical and electronic structures, and for the vibrational
properties of the systems we have investigated. The last Section is
devoted to our conclusions.

\section{Computational framework}
\label{sec:theo}

All the calculations have been performed within density functional
theory (DFT) using the generalized gradient approximation (GGA) with
the Perdew-Burke-Ernzerhof (PBE) exchange-correlation
functional.\cite{PBE} Vibrational properties have been calculated
using density-functional perturbation theory
(DFPT).\cite{baro87,gian91,dalc97,baro01} We have used the
pseudopotential method with ultra-soft pseudopotentials \cite{vand90}
and plane-wave basis sets up to a kinetic-energy cutoff of 27 Ry (216
Ry for the charge-density cutoff). Details about the Ag and O
pseudopotentials are reported in Ref.~\onlinecite{cipr02}.  Brillouin
zone (BZ) integrations have been performed with the Gaussian-spreading
special-point technique\cite{monk76,meth89} with a smearing parameter
of 0.03 Ry. All the calculations have been performed using the {\tt
PWscf} and {\tt PHONON} packages,\cite{PWSCF_WEB} while
molecular graphics has been generated with the {\tt XCRYSDEN}
package.\cite{XCRYSDEN}

Surfaces are modeled with periodic super-cells. For the Ag(410) and
Ag(210) surfaces we have used slabs of 20 (410) layers and 14 (210)
layers, respectively. For the Ag(100) surface and the (2$\times$1)
added-row reconstructed Ag(110) surface [Ag(110)p(2$\times$1)O] we
have used slabs of 7 layers. Adjacent slabs are separated by a vacuum
region of at least 16 a.u. and O atoms are adsorbed on both sides of
the slabs. All the structures have been fully relaxed until the
Hellmann-Feynman forces were smaller than $10^{-3}$ Ry/a.u. Atomic
oxygen adsorption on Ag(410) and Ag(210) has been modeled by
(1$\times$1) and (2$\times$1) surface super-cells. For the Ag(410)
surface the BZ integrations have been performed using a
(8$\times$4$\times$1) and a (4$\times$4$\times$1) uniform shifted
$k$-mesh\cite{kmesh} for the (1$\times$1) and (2$\times$1) structures,
respectively, while for Ag(210) ($8\times7\times1$) and
($4\times7\times1$) meshes have been used for (1$\times$1) and
(2$\times$1) structures, respectively.  For the c(2$\times$2) cell on
Ag(100) we have used a ($9\times9\times1$) mesh, while for the
Ag(110)p(2$\times$1)O surface we have used a ($6\times9\times1$)
mesh. Spin polarization effects have been neglected after checking
that the interaction between O and Ag substrate results in a
negligible spin moment on the O atom.

The chemisorption energies, $E_{\rm chem}$, are referred to the clean
Ag($n$10), $n=4$ or $2$, surface and the isolated oxygen molecule:
\begin{equation}
E_{\rm chem} = (E_{\rm O/Ag} - E_{\rm Ag} - N_{\rm O}(E_{\rm O_2}/2))/
N_{\rm O} ,
\end{equation}
where the total energy of the adsorbate--substrate system, of the
clean surface, of the isolated O$_2$ molecule, and the number of
adsorbed oxygen atoms are represented by $E_{\rm O/Ag}$, $E_{\rm Ag}$,
$E_{\rm O_2}$ and $N_{\rm O}$, respectively. With this definition,
stable adsorbates have negative chemisorption energies. 
Considering the errors in the structural relaxation and the
energy convergence with respect to the plane wave cutoff energy and
the k-point density, we estimate an overall numerical accuracy for the
chemisorption energies of about 30 meV.

The PBE GGA functional used in the present work slightly overestimates
the binding energy of the O$_2$ molecule: 5.50 eV, as calculated using
spin-polarized GGA with a 15 \AA-wide cubic box, while the
experimental value is 5.23 eV.\cite{O2} Molecular overbinding is a
common drawback of current implementations of DFT. This fact, of
course, does not affect the conclusions of the present work which
mainly concerns the {\it relative} stability of different adsorption
configurations.
    
Vibrational frequencies are calculated by diagonalizing the ${\bf
q}=0$ dynamical matrices of large slabs containing more that 50 atomic
layers.  The common procedure for calculating the dynamical matrices
of such large slabs is to patch the dynamical matrices obtained for
much smaller slabs (in fact, the same which were used to determine the
structural properties, as described above) with the interatomic force
constants calculated for the bulk metal.\cite{baro01} The dynamical
matrices of the Ag bulk have been calculated on a 4$\times$4$\times$4
grid of {\bf q}-points in the BZ and the interatomic force constants
have been obtained by a Fourier transform.\cite{baro01} For the thin
slabs, we have computed only the parts of the dynamical matrices which
refer to the displacements of the adsorbates and of the uppermost Ag
layers. Diagonalizing the dynamical matrix of the extended slab we
have calculated the vibrational modes of the systems and we have
identified the adsorbate-substrate modes from the displacements
eigenvectors. The estimated numerical accuracy of the vibrational
frequencies so obtained is of about 1 meV.

\section{Results}
\label{sec:res}

%
%

\subsection{Energetics}

\begin{figure}
  \centering
  \begin{tabular}{cc}
    ~~~~~Ag(410) & ~~~~~Ag(210)\\
    \includegraphics*[width=5.0cm]{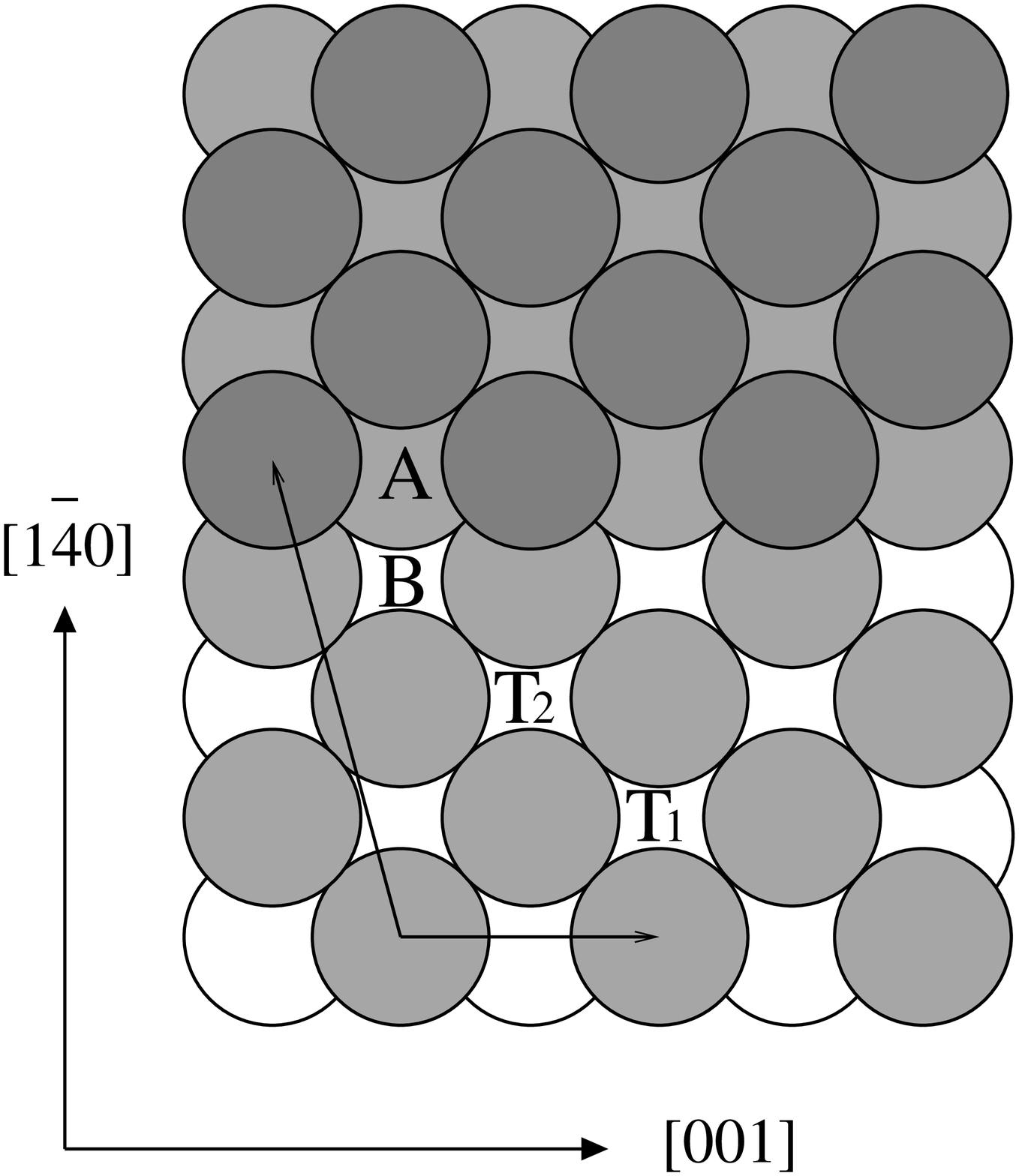}&
    \includegraphics*[width=5.0cm]{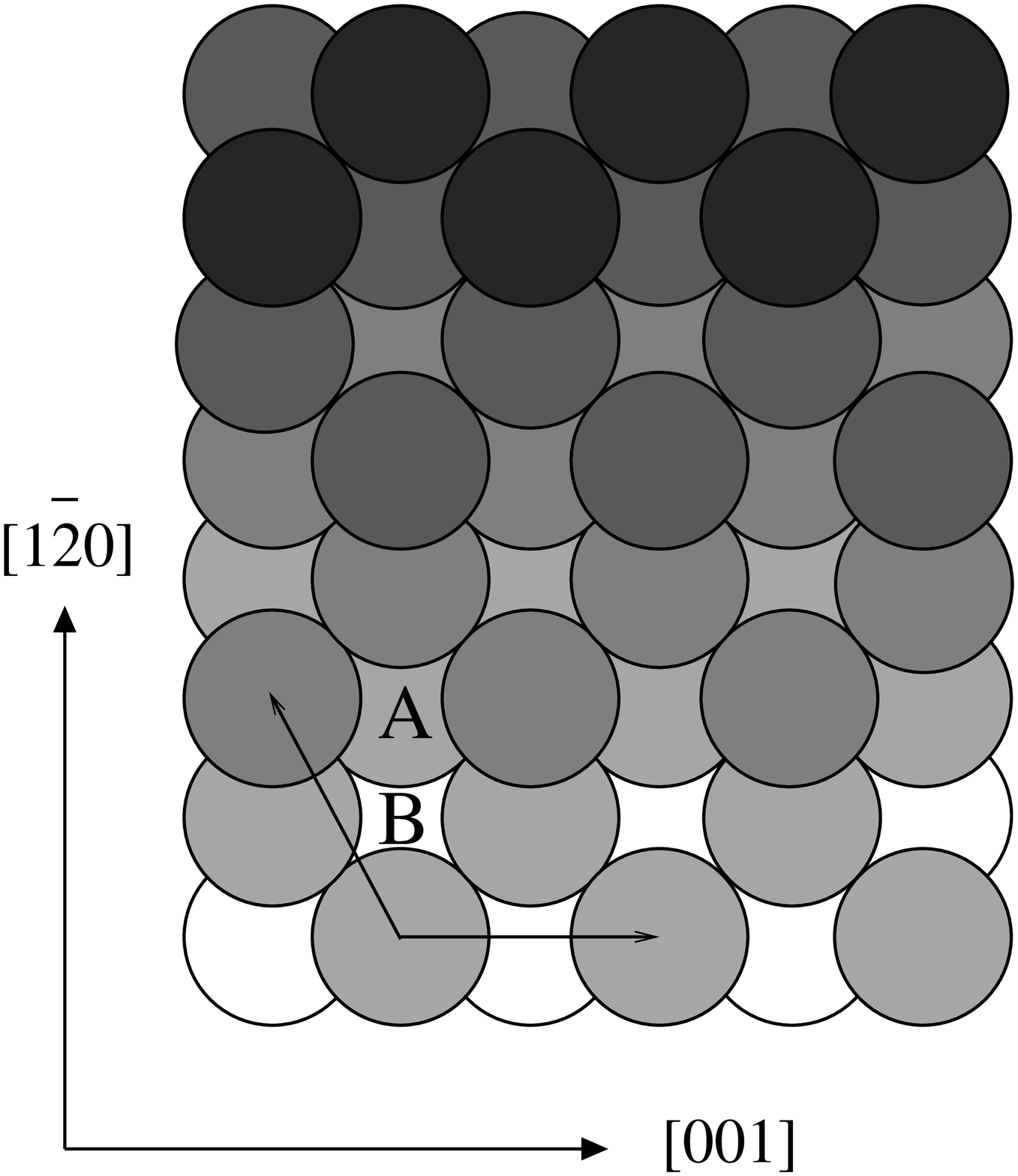}\\
  \end{tabular}
  \caption{
    Schematic representation of the Ag(410) (left) and Ag(210) surface
    (right). The labels A, B, T$_1$ and T$_2$ indicate different
    adsorption sites: A, above the step; B, below the step; T$_1$ and
    T$_2$, hollow sites at terrace.  The surface unit-cell vectors,
    and the [$001$] and [$1\bar{n}0$], $n=4,2$, crystal axes are also
    shown.
  } 
  \label{fig:surf}
\end{figure}

In Fig. \ref{fig:surf} we display the four adsorption sites that we
have considered in the present work: site A with O just above the step
and coordinated with three Ag surface atoms; sites T$_1$ and T$_2$
with the adatom in the hollow sites on the (100) terrace; site B where
O adatom lies in the hollow site just below the step. Note that no
terrace T$_1$ and T$_2$ sites are available on the Ag(210).  Top and
bridge sites are not considered here since we have found that they are
much less stable than hollow sites, as on Ag(100)
surface.\cite{cipr02}


\subsubsection{O/Ag(410)}

\begin{figure*}
  \centering
  \begin{tabular}{cccc}
    A & T$_1$ & T$_2$ & B \\
    \includegraphics[width=0.155\textwidth]{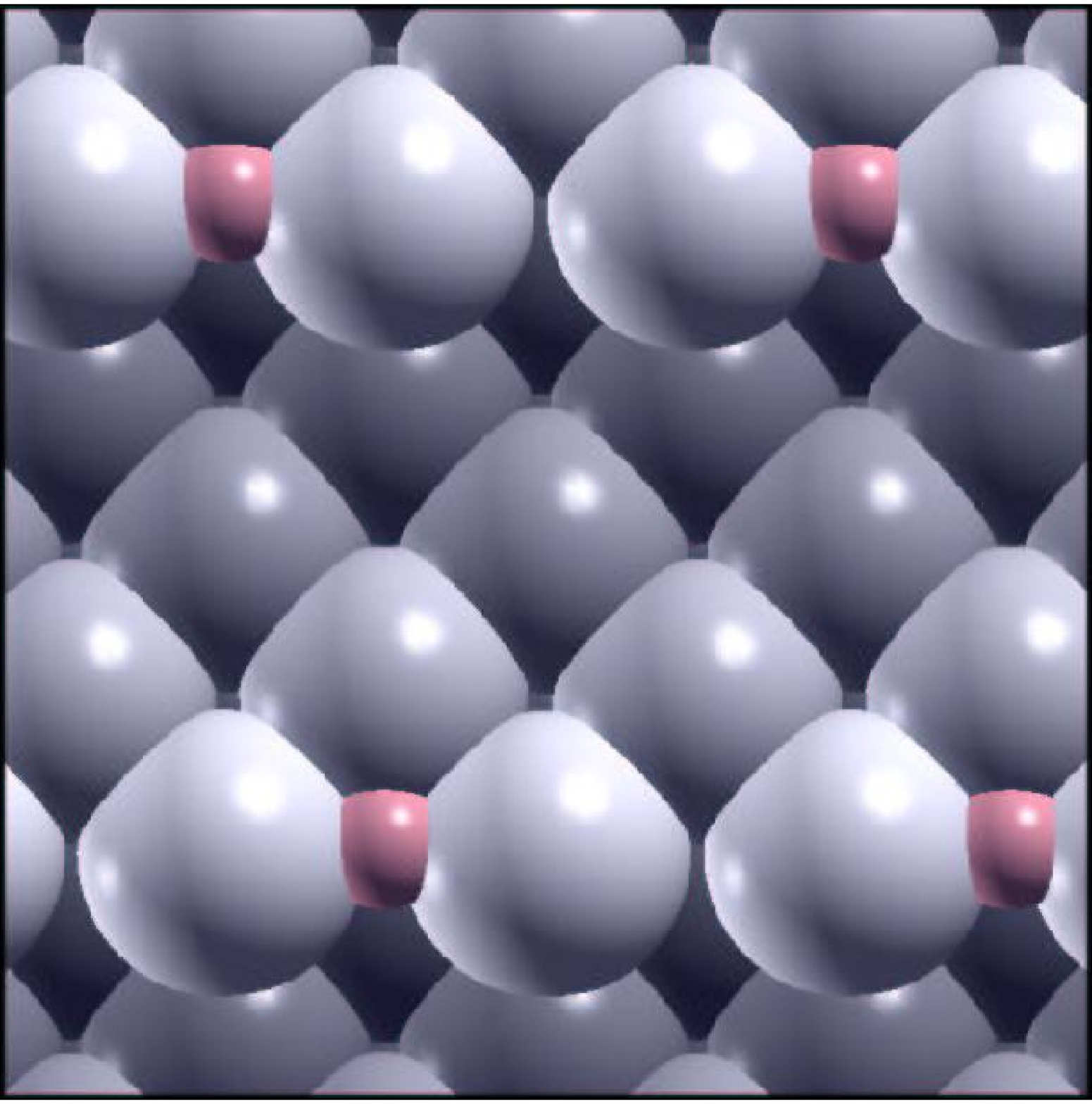}&
    \includegraphics[width=0.155\textwidth]{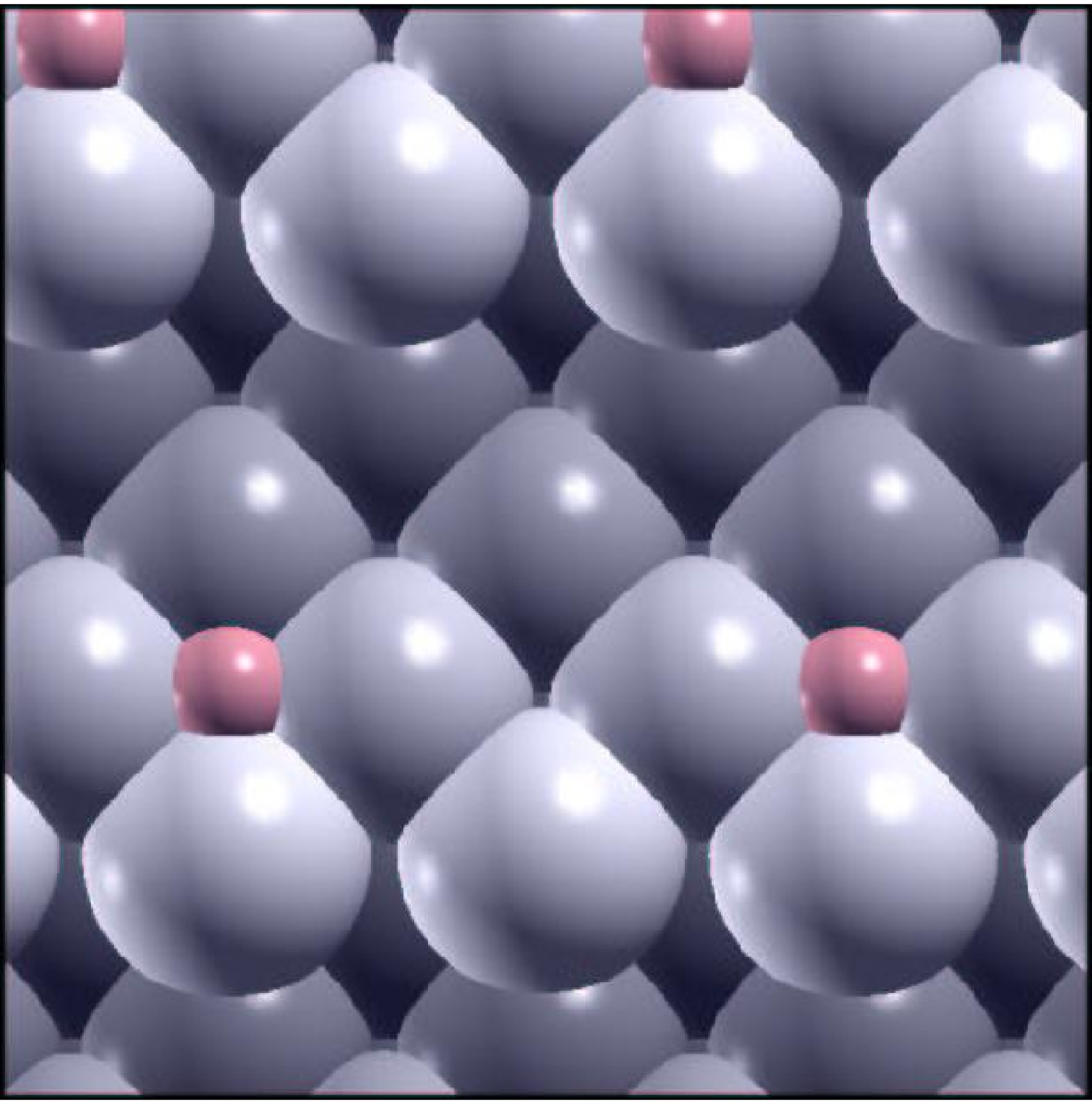}&
    \includegraphics[width=0.155\textwidth]{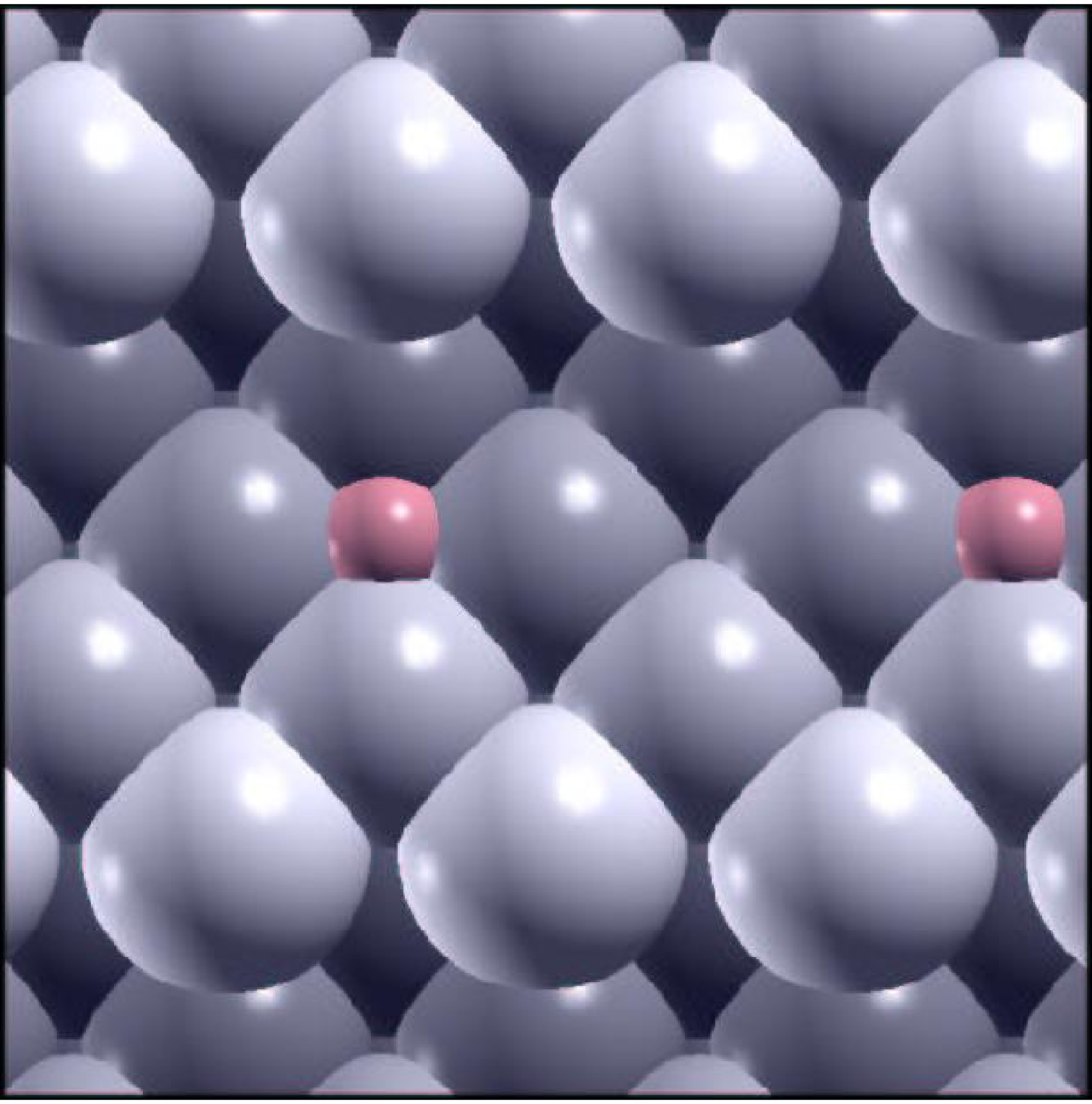}&
    \includegraphics[width=0.155\textwidth]{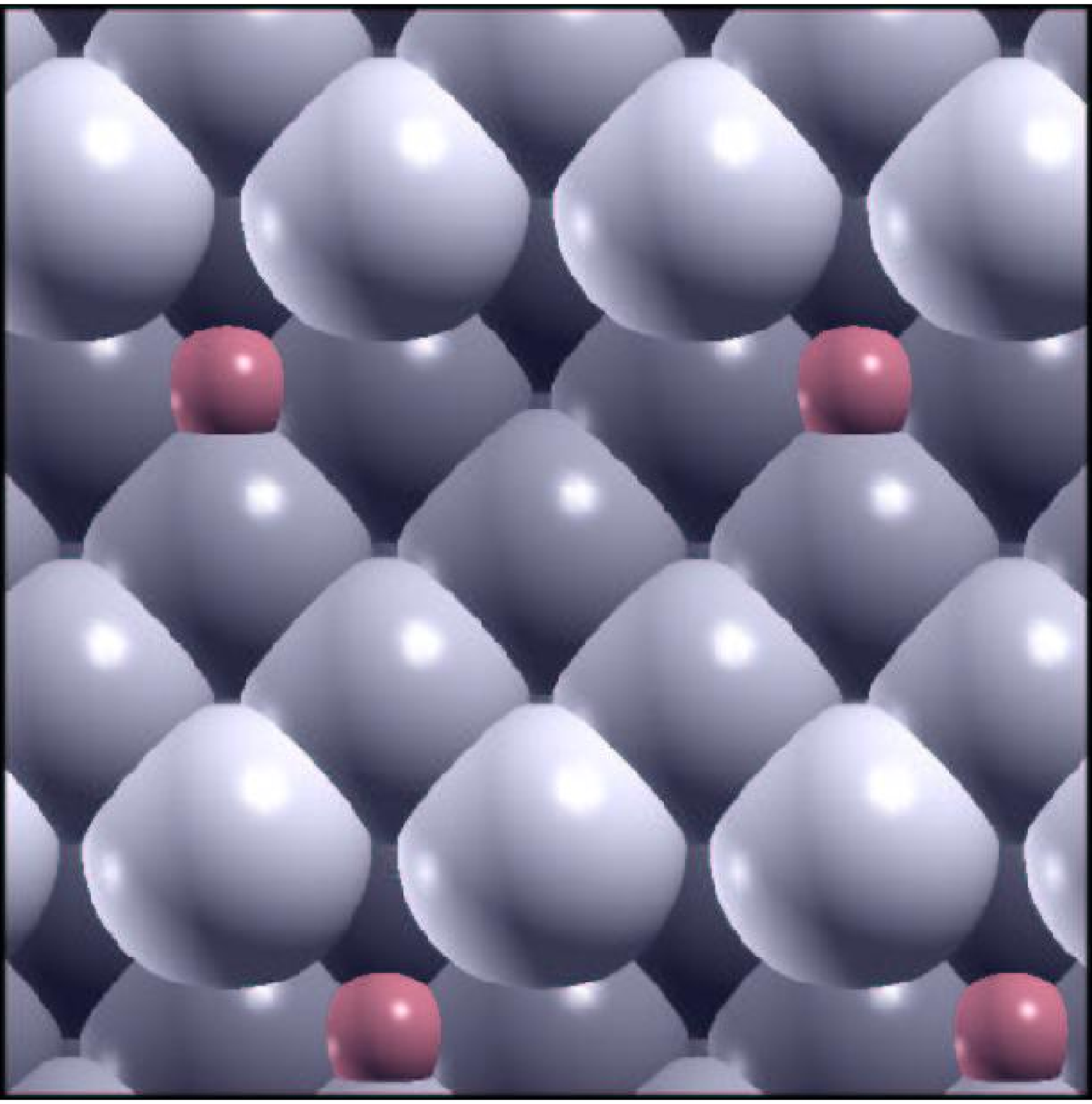}\\[1em]
    A--A & T$_1$--T$_1$ & T$_2$--T$_2$ & B--B \\
    \includegraphics[width=0.155\textwidth]{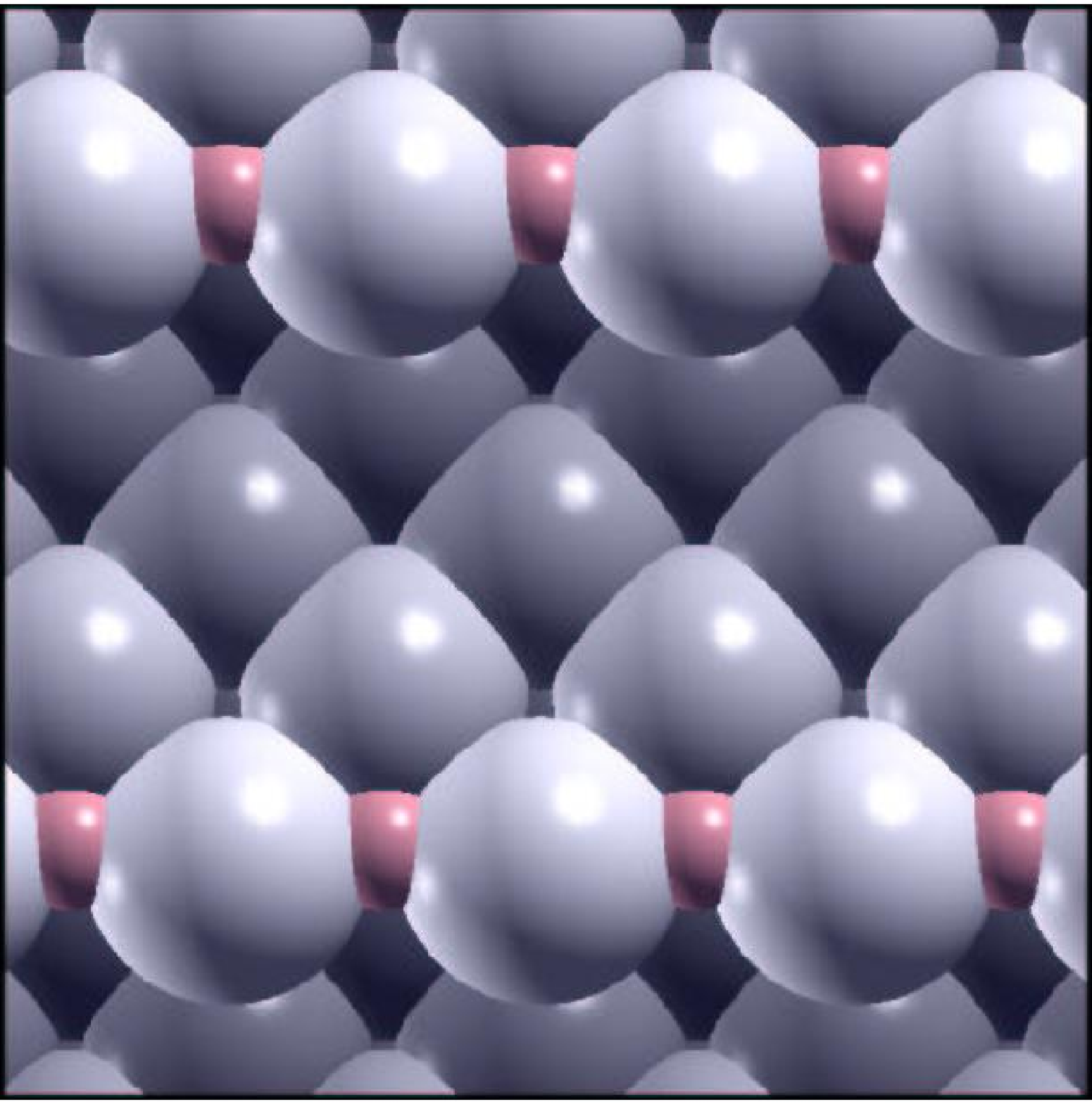}&
    \includegraphics[width=0.155\textwidth]{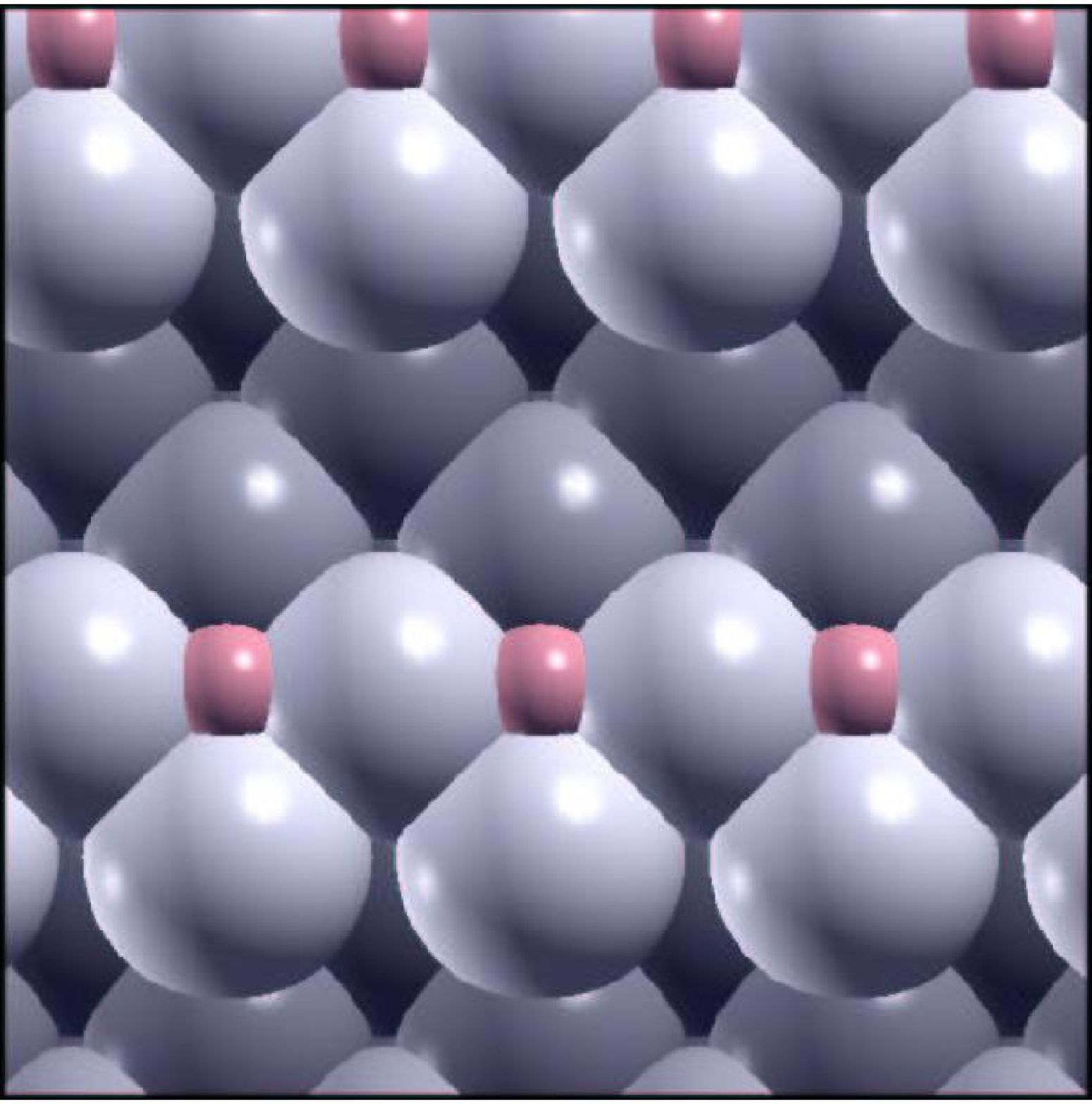}&
    \includegraphics[width=0.155\textwidth]{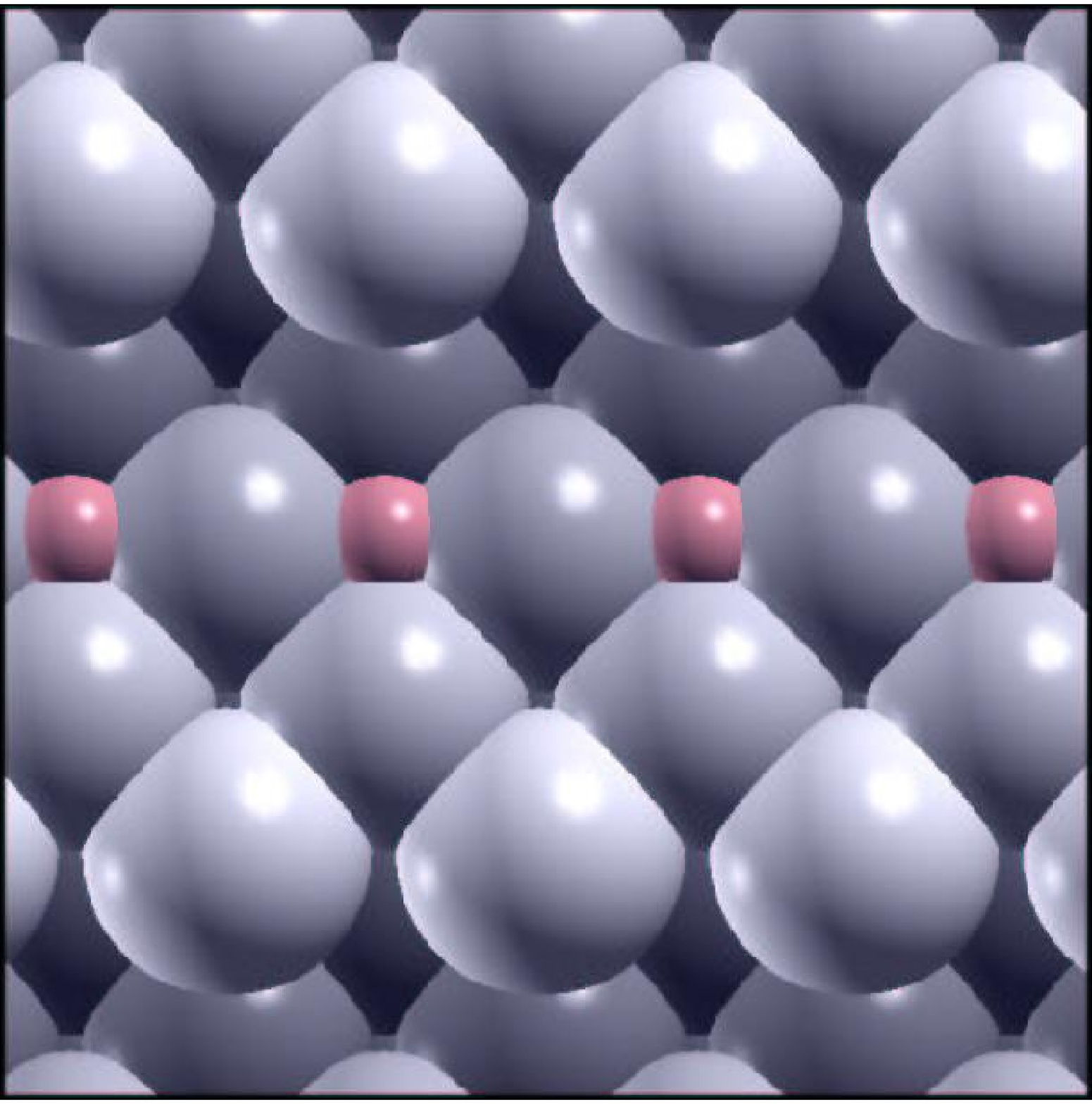}&
    \includegraphics[width=0.155\textwidth]{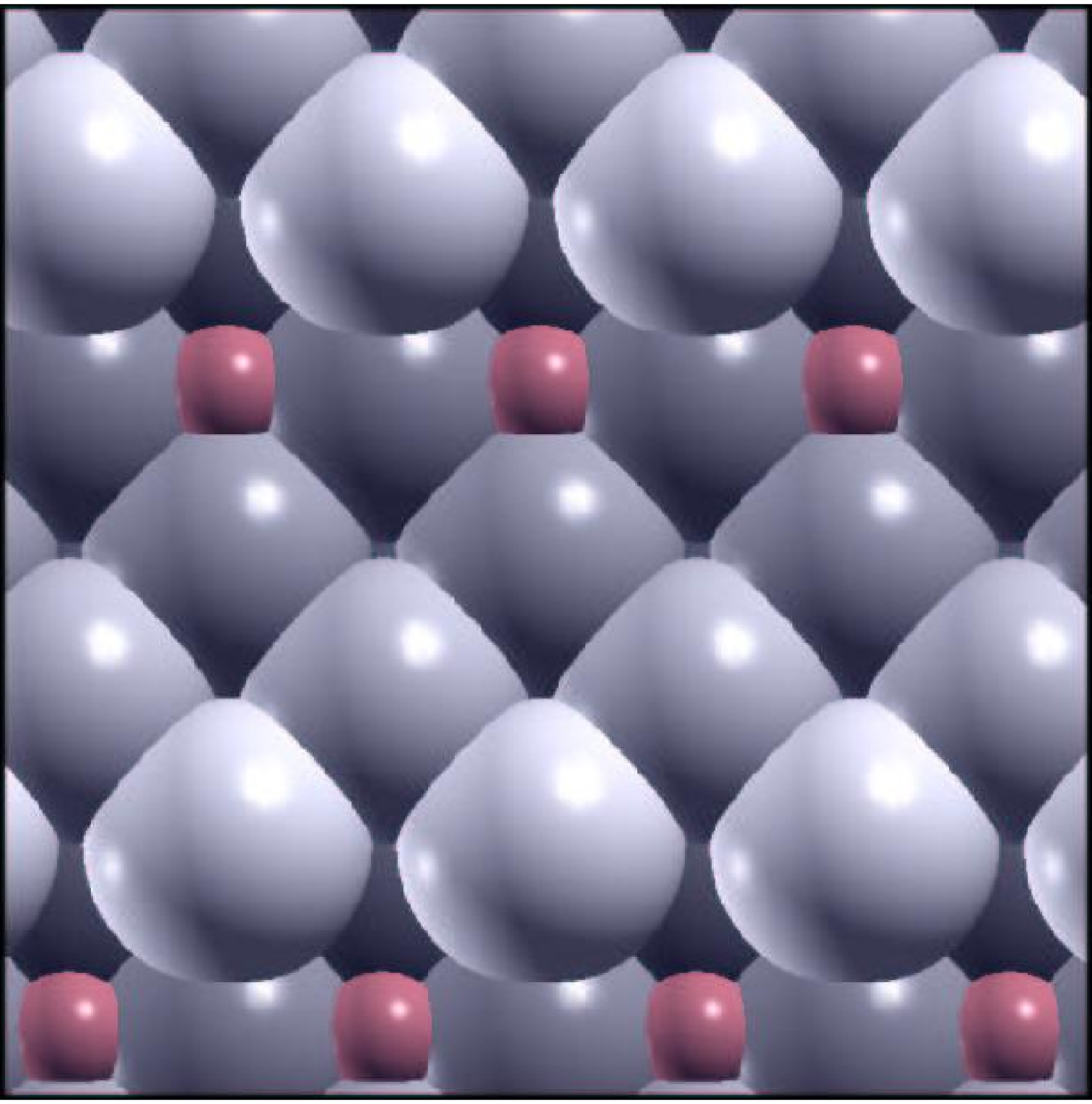}\\[1em]
  \end{tabular}
  \begin{tabular}[b]{cccccc}
    A--T$_2$ & T$_1$--B & A--T$_1$ & T$_1$--T$_2$ & T$_2$--B & A--B\\
    \includegraphics[width=0.155\textwidth]{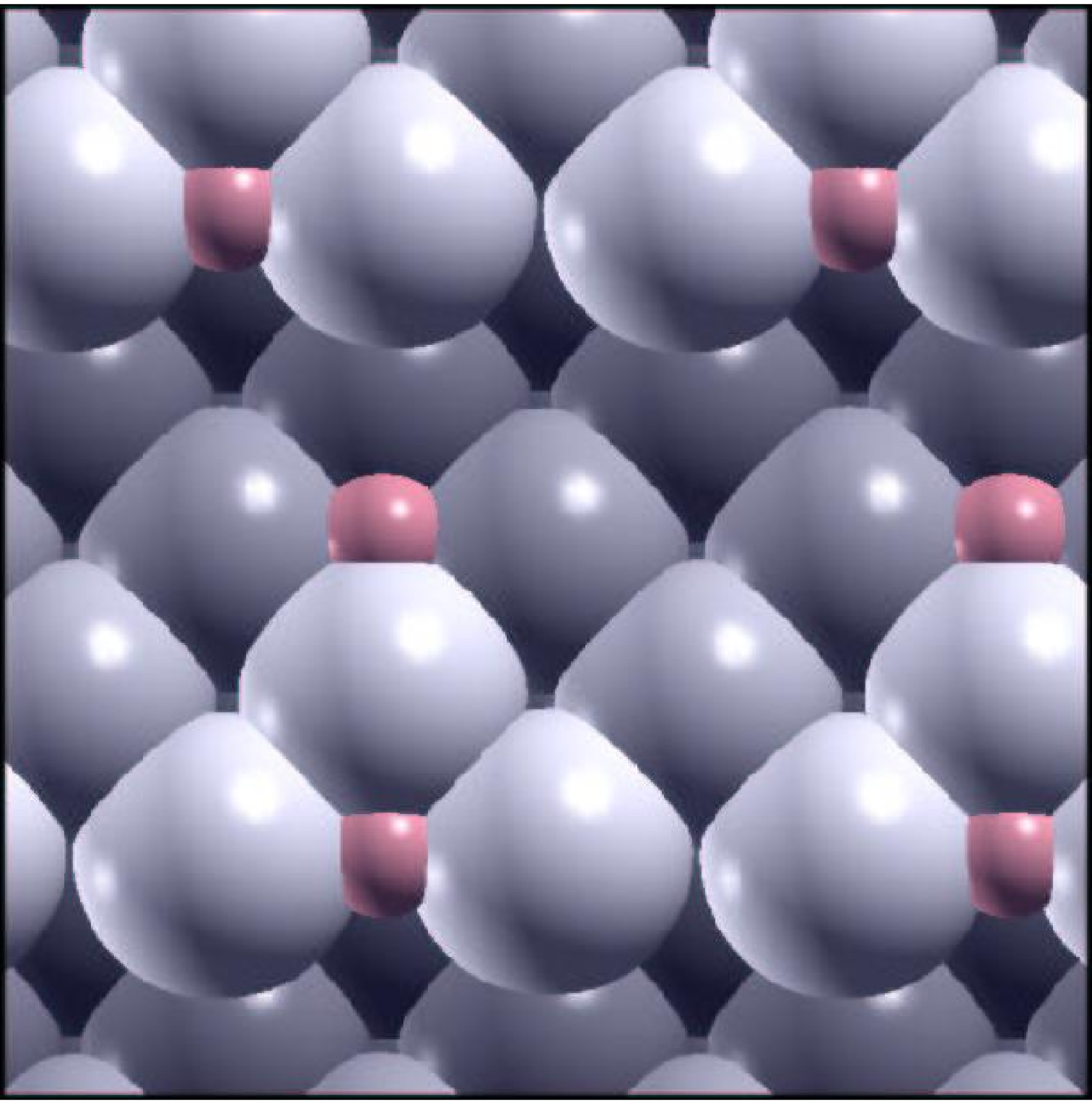}&
    \includegraphics[width=0.155\textwidth]{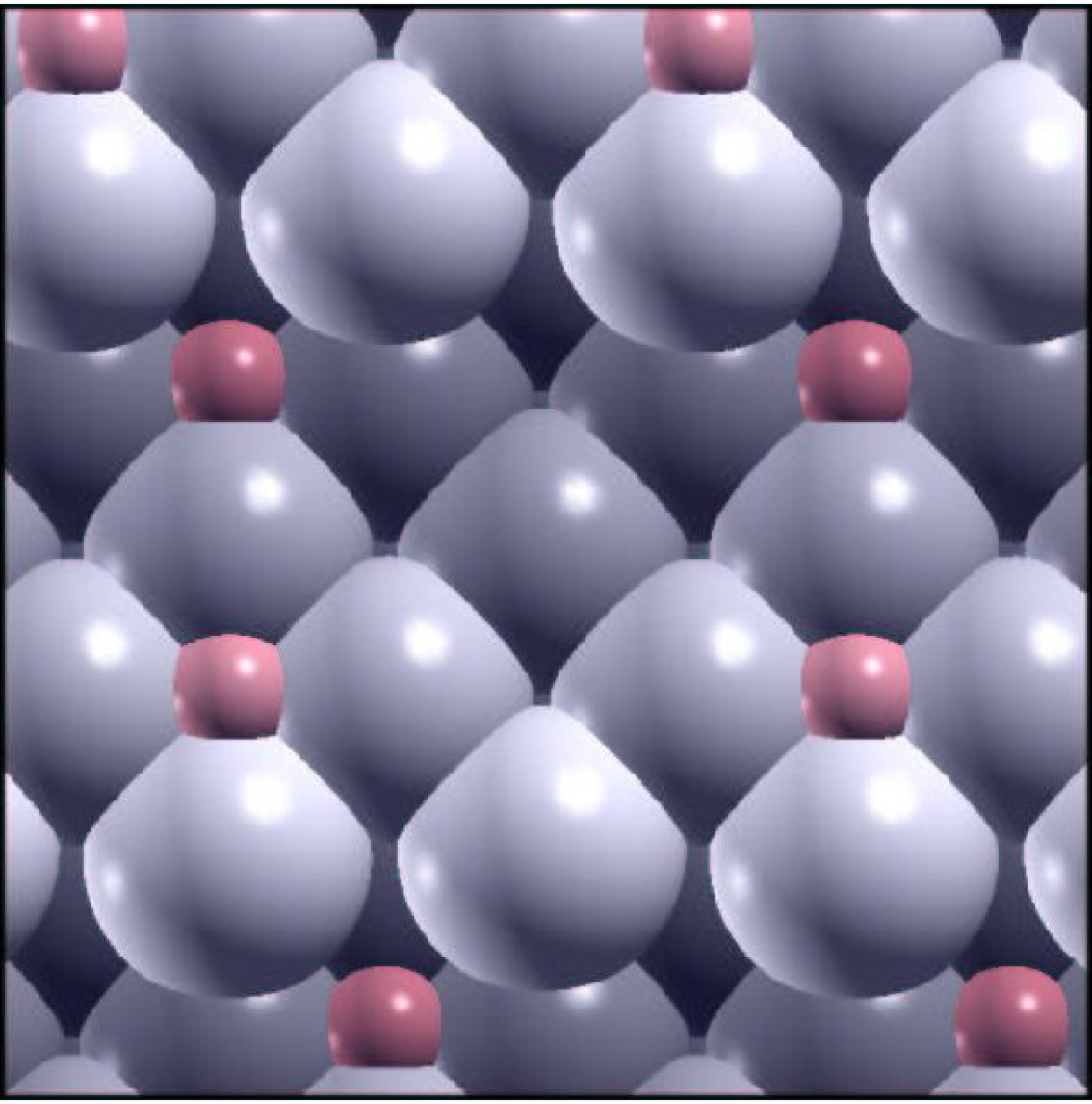}&
    \includegraphics[width=0.155\textwidth]{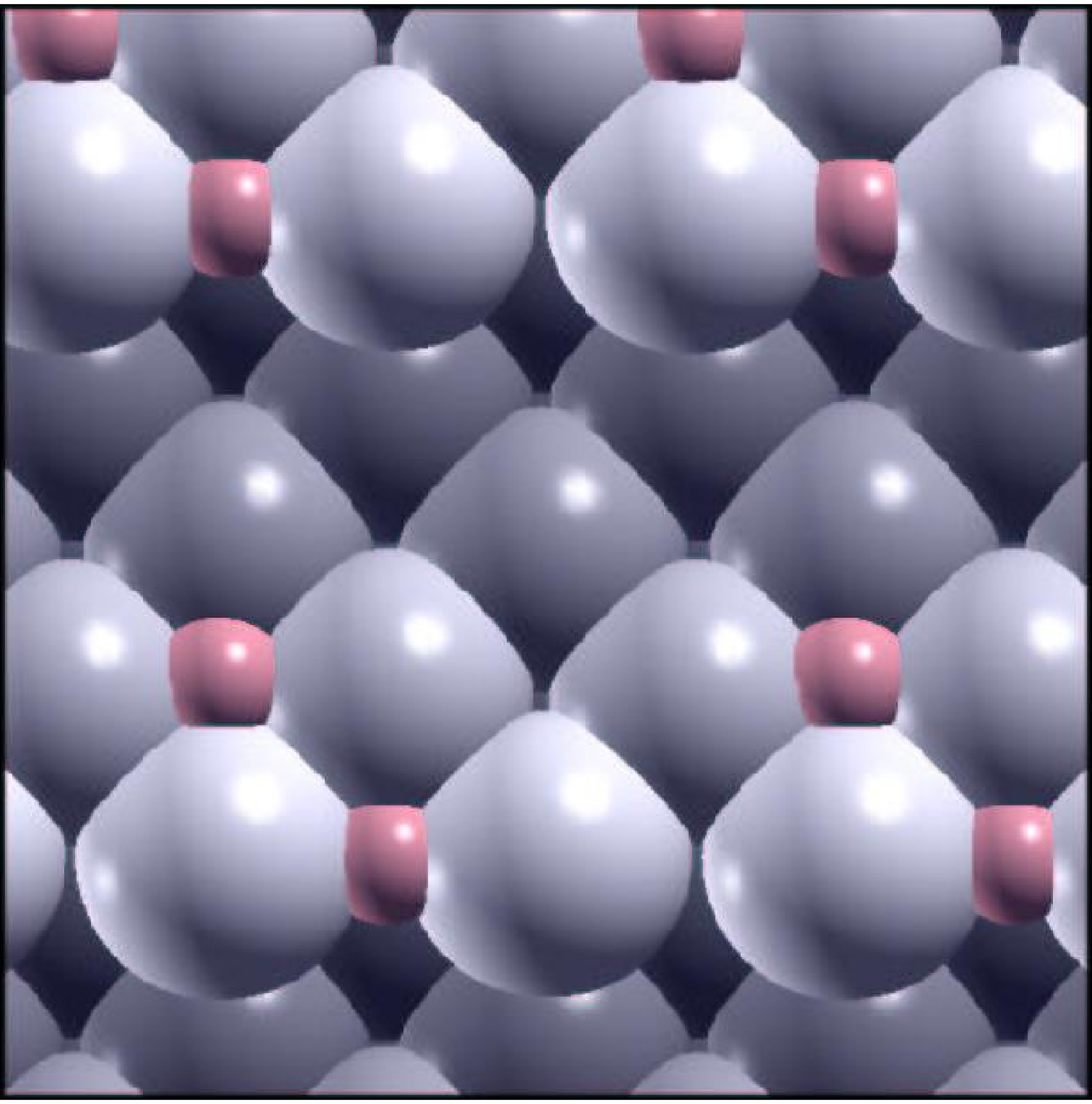}&
    \includegraphics[width=0.155\textwidth]{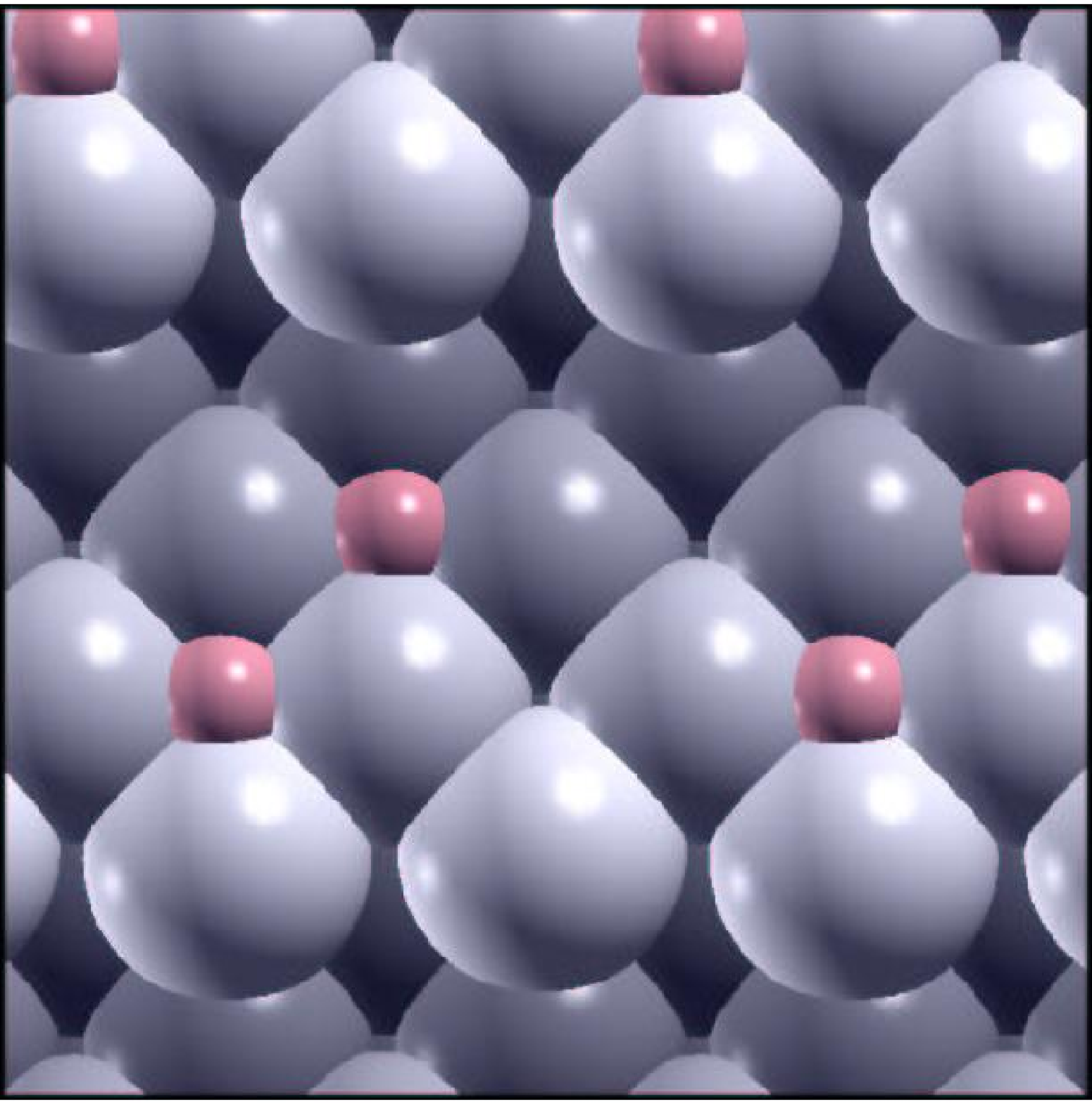}&
    \includegraphics[width=0.155\textwidth]{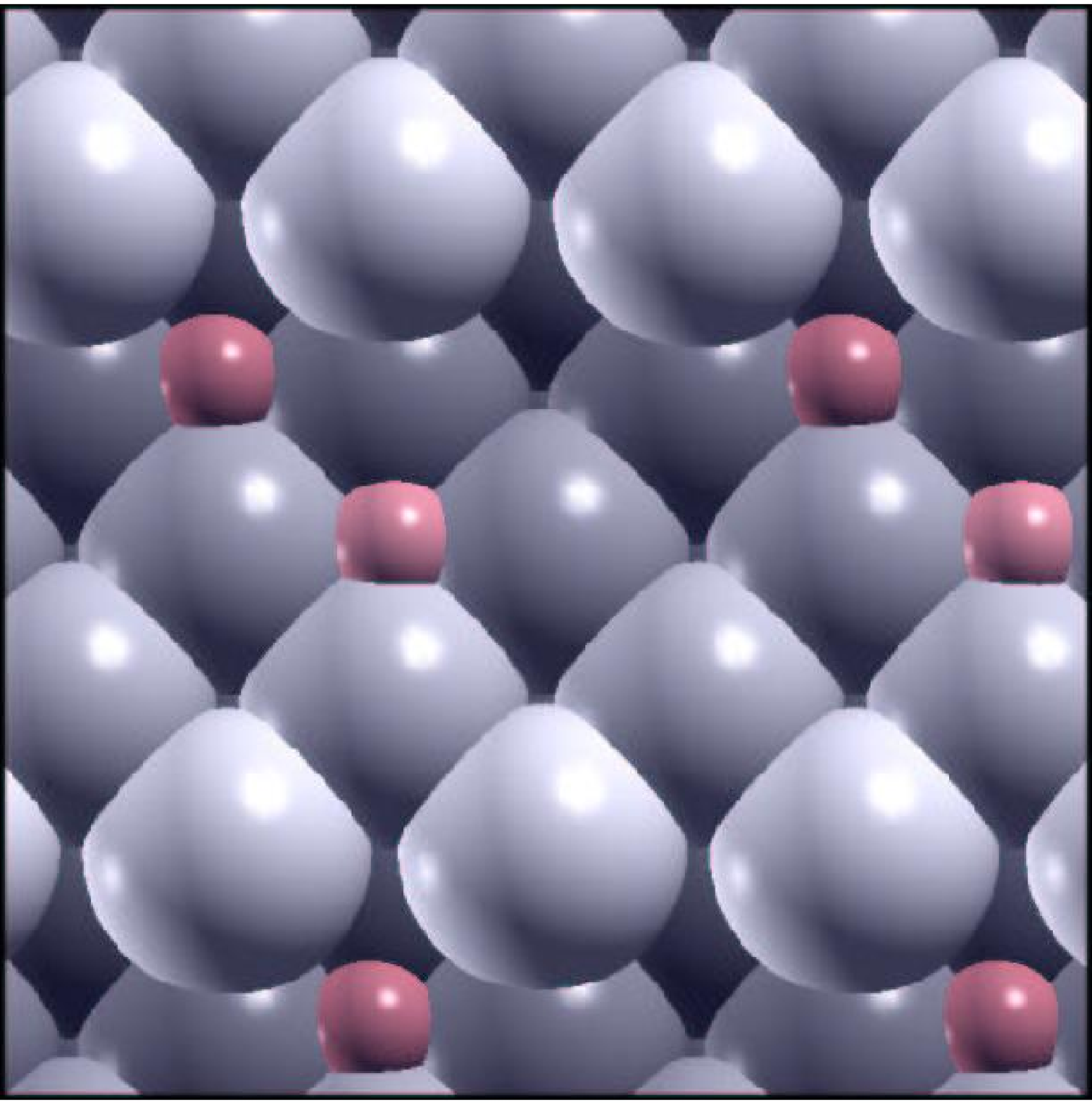}&
    \includegraphics[width=0.155\textwidth]{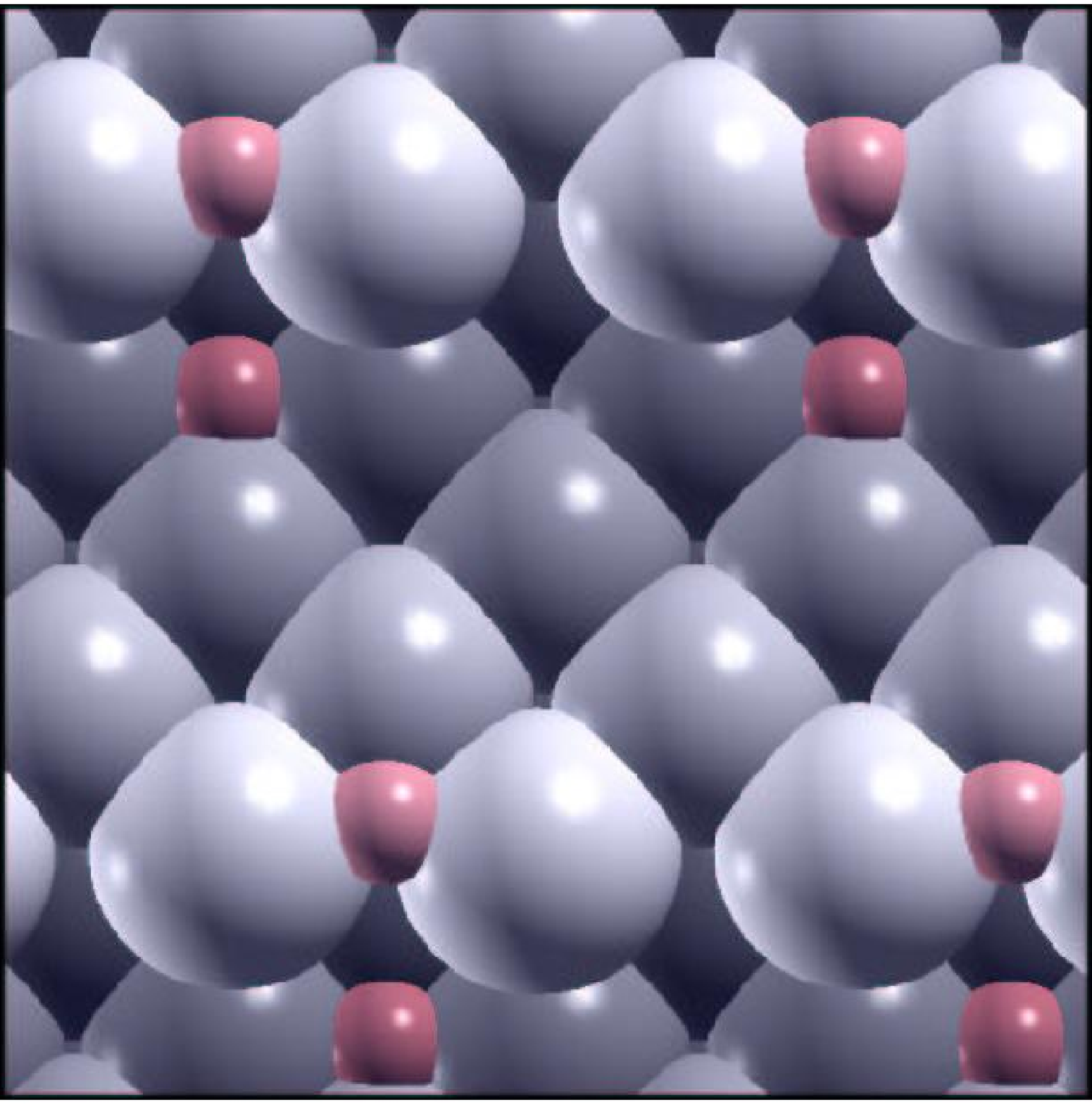}\\
  \end{tabular}
  \caption{(Color online) Top views of optimized S$_1$, S$_1$--S$_1$,
    and S$_1$--S$_2$  O/Ag(410) structures. Grey (red) balls represent
    silver (oxygen) atoms.}
  \label{fig:configs}
\end{figure*}

Let us first consider the adsorption of atomic oxygen on the Ag(410)
surface.  In Table \ref{tab:1} we present the chemisorption energies
corresponding to different adsorption configurations (see
Fig.~\ref{fig:configs}).  At low coverage ($\Theta =$ 1/8
ML)\cite{cover} where the adatoms stay far apart from each other, the
hollow sites on terraces (T$_1$ and T$_2$) and the step-edge site (A)
are almost degenerate in energy, with terrace sites being only slightly
more favored than site A, while site B is much less stable.  In these
configurations the distances between the O atoms are quite large (8.32
{\AA}) and this suggests a negligible interaction between the adatoms,
as confirmed by test calculations done at lower coverages in which the
O$-$O distance is increased.  At higher coverage ($\Theta =$ 1/4 ML)
we have examined various possible geometries which we indicate with
the notation `S$_1$--S$_2$', where the S's stand for two near A, B,
T$_1$ and T$_2$ sites being occupied by two O adatoms in a
(2$\times$1) super-cell. The S$_1$--S$_1$ arrangements correspond to a
(1$\times$1) periodicity where O adatoms form rows parallel to the
steps. Those configurations in which the separation between the
adatoms is larger than 5 {\AA} are not considered since we expect that
their adsorption energies should be very similar to those at low
coverage. Indeed, as we will show, already for O--O distances as small
as 4 {\AA} the interaction between the adsorbates has small effects on
the chemisorption energies.  Our results show that the S$_1$--S$_1$
configurations with both the adatoms on the terrace or above the step
and the A--T$_2$ arrangement are the most stable adsorption
geometries. As in the case of low coverage the differences in
chemisorption energy between these sites are quite small, but in this
case we find that oxygen adatoms slightly prefer to decorate the step
edges.  One expects that occupying adjacent sites should be
unfavorable because of the electrostatic repulsion between negatively
charged oxygen adatoms. Indeed, for example, Feibelman and
co-workers\cite{feib96} have shown that O adatoms on Pt(211) vicinal
surface decorate the steps in a (2$\times$1) adsorption geometry while
the (1$\times$1) structure---where the distance between adjacent
oxygens is about 2.8 {\AA}---is significantly less stable. Our results
indicate that this effect on Ag(410) surface is important only when
nearest neighbor sites are occupied, i.e. when the distance between
the adatoms is less than 4 {\AA} (as in A--T$_1$, T$_1$--T$_2$,
T$_2$--B and A--B, configurations), while it is weak and uneffective
in the other geometries. Compare for example the A--T$_2$ and A--T$_1$
configurations which consist of one adatom sitting above the step
while the other sits at a terrace site. In the A--T$_2$ arrangement
the distance between the O atoms is 4.23 {\AA} and their chemisorption
energy ($-$0.80 eV) is quite close to the values found for oxygen
adsorption at sites A and T$_2$ at low coverage ($-$0.75 eV and
$-$0.83 eV), indicating a small adsorbate-adsorbate interaction. In
the A--T$_1$ configuration, instead, the O--O distance is only 3.26
{\AA} and the adsorption energy of the adatoms ($-$0.59 eV) is
strongly reduced with respect to when oxygen is adsorbed at sites A
and T$_1$ at low coverage ($-$0.75 eV and $-$0.80 eV). Moreover, it is
interesting to observe that in the A--T$_1$ geometry the electrostatic
repulsion strongly affects also the locations of the adatoms. Indeed
we find that the adsorbates are slightly displaced from the hollow
positions and that their distance, 3.26 \AA, is considerably larger
than the separation between two nearest hollow sites of the clean
surface (about 2.9 {\AA}). This is a clear indication that the two
adatoms repel each other.


\subsubsection{O/Ag(210)}

Similar calculations have been performed also for the Ag(210) surface.
The two surfaces differ in the width of the terrace (see Fig.
\ref{fig:surf}). No terrace hollow sites are available on the Ag(210)
surface. In Table \ref{tab:1b} we report the chemisorption energies
for O adsorbed in various configurations on this surface.  These
results display the same trend as obtained for Ag(410), although the
chemisorption energies are somewhat smaller here. Due to the lack of
stable adsorption sites on the terraces, the formation of O--Ag--O
rows at the step edges (i.e.  the A--A configuration) is by far
energetically favored on this surface.  The second most stable
configuration at higher coverage ($\Theta=1/2$ ML) is therefore A--B,
which is 0.24 eV less bound than A--A. In contrast, on Ag(410) several
other configurations that involve terrace sites are more stable than
A--B and much closer in energy to the A--A configuration. Comparing
the chemisorption energies of various configurations, irrespective of
the O coverage, we find that the difference between the most favored
A--A configuration and the second most stable one is 0.03 and 0.12~eV
for Ag(410) and Ag(210) surfaces, respectively.


\subsubsection{The stability of O--Ag--O rows}

\begin{figure}
  \centerline{\includegraphics*[width=8.0cm]{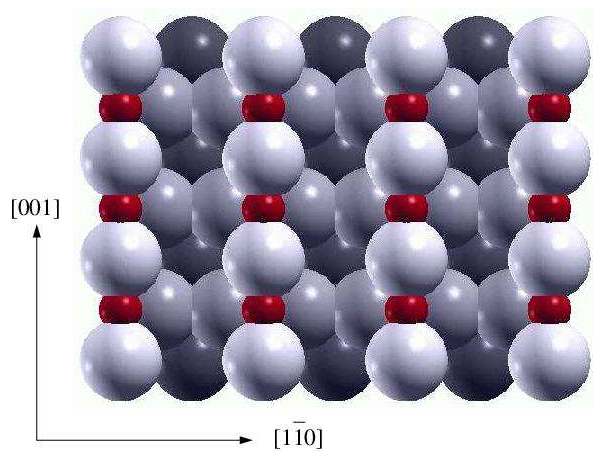}}
  \caption{(Color online) Top view of the Ag(110)p(2$\times$1)O surface.
    Grey (red) balls represent silver (oxygen) atoms.}
  \label{fig:added}
\end{figure}

One of the most peculiar results of our calculations is the increase
of the stability of oxygen adatoms when they decorate the step edge in
a (1$\times$1) geometry.  We believe that the mechanism responsible
for this behavior could be related to the arrangement of O adatoms to
form O--Ag--O chains at the upper side of the (110) steps.  Indeed,
the A--A adsorption configuration is similar to that occurring in the
reconstruction of the Ag(110) surface upon oxygen adsorption, in which
O atoms align in added-rows along the [001] direction, thus forming
stable O--Ag--O structures\cite{cour97,schi93} (the Ag(110)p(2$\times$1)O
surface is shown in Fig. \ref{fig:added}). As we have pointed out
in the introduction, this similarity between the two structures was
first suggested from a comparison of the vibrational properties of the
two surfaces.\cite{savi02}

Before analyzing in detail the structural and electronic features
related to the formation of these chains, it is interesting to
consider the effects of the substrate relaxation on the chemisorption
energies.  For this reason we have repeated some calculations for
O/Ag(410) keeping the substrate fixed in the optimized clean surface
geometry and allowing the adatoms to relax (see values in parentheses
in Table \ref{tab:1}).  Our results show that at low coverage the
T$_2$ site is more stable than A site even if substrate relaxation is
not considered.  On the other hand, in the (1$\times$1) periodicity O
adatoms in A--A and T$_2$--T$_2$ configurations have the same
chemisorption energy on a fixed substrate. This result indicates that
the enhanced stability of atomic oxygen on the step in the
(1$\times$1) geometry is mainly determined by the energy gain due to
surface relaxation allowed by the presence of the steps.

%
%

\subsection{Geometrical structure}
\begin{figure}
  \centering
  \includegraphics[width=8.0cm]{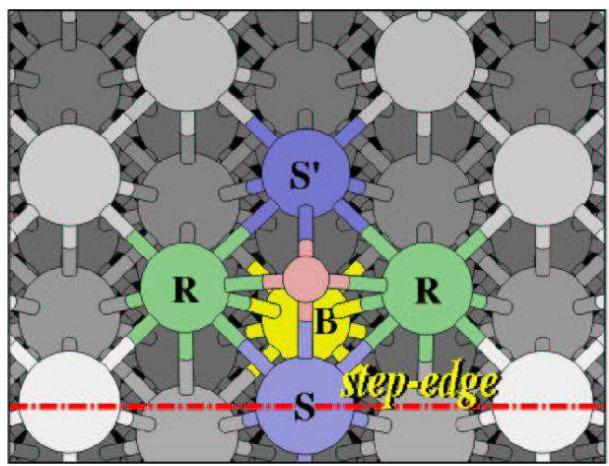}
  \caption{(Color online) A schematic definition of Ag$^{\rm R}$,
    Ag$^{\rm S}$, and Ag$^{\rm B}$ atoms, which are labelled as R, S
    (or S'), and B, respectively. Larger balls are Ag atoms, while
    smaller red ball is O adatom. The step-edge is marked with
    dot-dashed line. The Ag$^{\rm R}$ atoms are lying in the row of
    oxygen adatoms, while Ag$^{\rm S}$ atoms are the other nearest
    surface silver atoms. The Ag$^{\rm B}$ silver atom is just below
    the O adatom.  }
  \label{fig:legend}
\end{figure}

In order to illustrate the characteristic features of the O--Ag--O
chain formation at the step edges, we discuss now the effects of
oxygen adsorption on the geometry of the surface. To facilitate the
discussion, we first define a few labels for silver atoms---Ag$^{\rm
R}$, Ag$^{\rm S}$, and Ag$^{\rm B}$---as in
Fig. \ref{fig:legend}.\cite{labels-note}


\subsubsection{O/Ag(410)}
In table \ref{tab:2} we report the optimized structural parameters
describing the oxygen adsorption on Ag(410) at $\Theta =$ 1/8 ML and
at $\Theta =$ 1/4 ML for the S$_1$--S$_1$ configurations. First of all
it is interesting to observe the dependence of the bond lengths
between the adsorbate and the nearest metal atoms on the number of
neighboring substrate atoms. As predicted by effective medium theory,
\cite{hamm97} the larger the number of nearest metal atoms the longer
the bond length. Indeed the results in Table \ref{tab:2} (observe in
particular $d_{\rm O-Ag^R}$ and $d_{\rm O-Ag^B}$) show that the
three-fold coordinated step-edge is the site in which O is closest to
the surrounding silver atoms; O in the two four-fold coordinated
terrace sites has similar structural features, while O---when sitting
in the site just below the step, close to four atoms of the terrace
and to two atoms of the step---is the furthest from neighboring metal
atoms.

\begin{figure}
  \centerline{\includegraphics*[width=8.0cm]{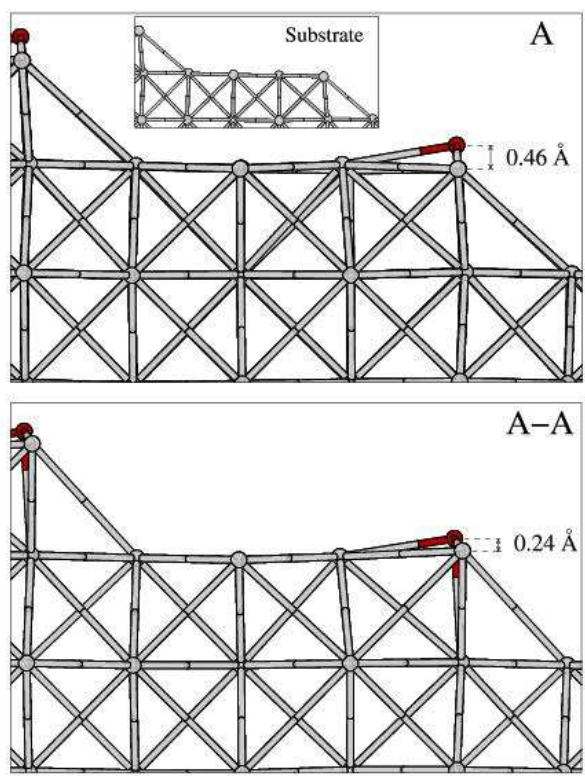}}
  \caption{(Color online) Side views of the first three layers of
    Ag(410) substrate with O adatoms on site A in the (2$\times$1) and
    (1$\times$1) geometries. Grey (red) balls represent silver
    (oxygen) atoms. The inset in the top panel shows the side view of
    the optimized clean Ag(410).}
  \label{fig:str1}
\end{figure}

We now focus on the structural features of oxygen adsorption at sites
A and T$_2$. In Fig.  \ref{fig:str1} we compare the structure of the
first three layers of the substrate for the A and A--A geometries. At
low coverage the main substrate distortions due to the presence of the
adatom are a shift of 0.18 {\AA} of the nearest Ag terrace atom
towards the center of the terrace (see $d_{{\rm Ag^{R}} -{\rm
Ag^{S'}}}$ in Table \ref{tab:3}) and an expansion of 0.16 {\AA} of the
distance between the two Ag atoms on the step near oxygen (see
$d_{{\rm Ag^{R}} -{\rm Ag^{R}}}$ in Table \ref{tab:3}). The result is
that the adatom sits in a site equidistant ($d_{{\rm O}-{\rm Ag}}$ =
2.21 {\AA}) from the three Ag surface atom surrounding it.  In the
A--A configuration the distortion has different features. In this case
symmetry does not allow the two Ag step atoms near oxygen to be
displaced parallel to the step, so that these atoms move only
outward. This is indicated by a marked expansion of the bond lengths
between the Ag step atom and the Ag atoms on terrace (by 0.25 {\AA})
and below the adatom (by 0.16 {\AA}).  In this case the oxygen is
closer to the Ag step atoms ($d_{{\rm O}-{\rm Ag^{\rm R}}}$ = 2.10
{\AA}) than to the nearest Ag atom on terrace ($d_{{\rm O} -{\rm
Ag^{S'}}}$ = 2.24 {\AA}). Note also that the angle, $\alpha$, between
the bonds of oxygen with the two Ag atoms of the step---which directly
measures the alignment of the Ag--O--Ag structures---is larger at high
than at low coverage (164$^\circ$ versus 156$^\circ$). Thus, in the
A--A geometry oxygen and silver atoms on the step tend to align
forming an almost straight O--Ag--O chain which is more stable than
other adsorption configurations. As we have already pointed out, the
O--Ag--O chains at the step edges have a similar structure as the
added rows in the reconstructed O/Ag(110) surface, so it is
interesting to compare the geometries of the two systems. In Table
\ref{tab:2} we report some predicted structural quantities for the
Ag(110)p(2$\times$1)O surface.  Our results are in good agreement with
experiments\cite{cane93} and previous DFT calculations.\cite{kata99}
As we can observe, the alignment of O and Ag atoms in the added rows
on Ag(110) is more pronounced than in the A--A configuration (compare,
for example, the angle $\alpha$ in the two geometries), but the
similarity between the two structures is clear.

\begin{figure}
   \centerline{\includegraphics*[width=8.0cm]{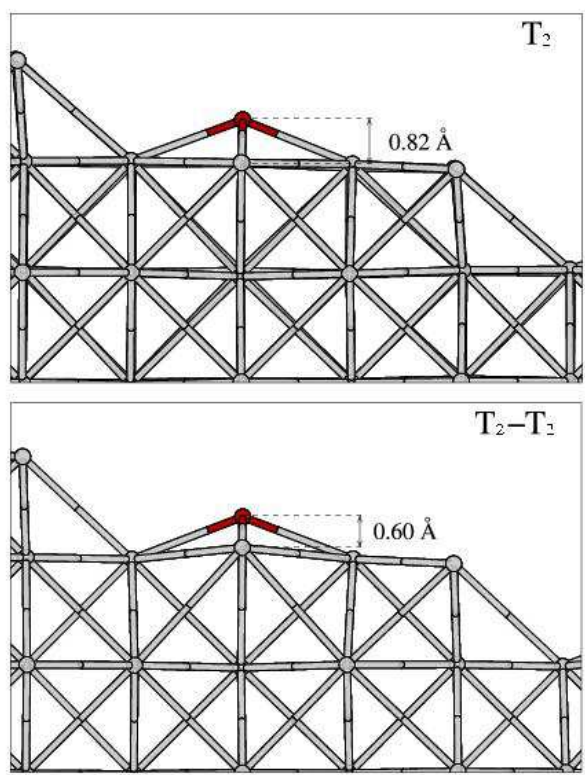}}
  \caption{(Color online) Same as in Fig. \ref{fig:str1}, but for T$_2$
    and T$_2$--T$_2$ configurations.\hfill}
  \label{fig:str2}
\end{figure}

Fig. \ref{fig:str2} shows the structures of T$_2$ and T$_2$--T$_2$
configurations. In these systems the height of the adatom with respect
to the terrace is larger than when the oxygen is chemisorbed on the
step. Also in this case in the (1$\times$1) geometry the silver atoms
in the row of oxygens move up towards adatoms, but here the alignment
between O and Ag atoms is less pronounced than when oxygen atoms are
on the step.


\subsubsection{O/Ag(210)}

\begin{figure*}
  \centering
  \begin{tabular}{c@{~~~~~~~~~~~~~}c}
    \includegraphics*[width=8.0cm]{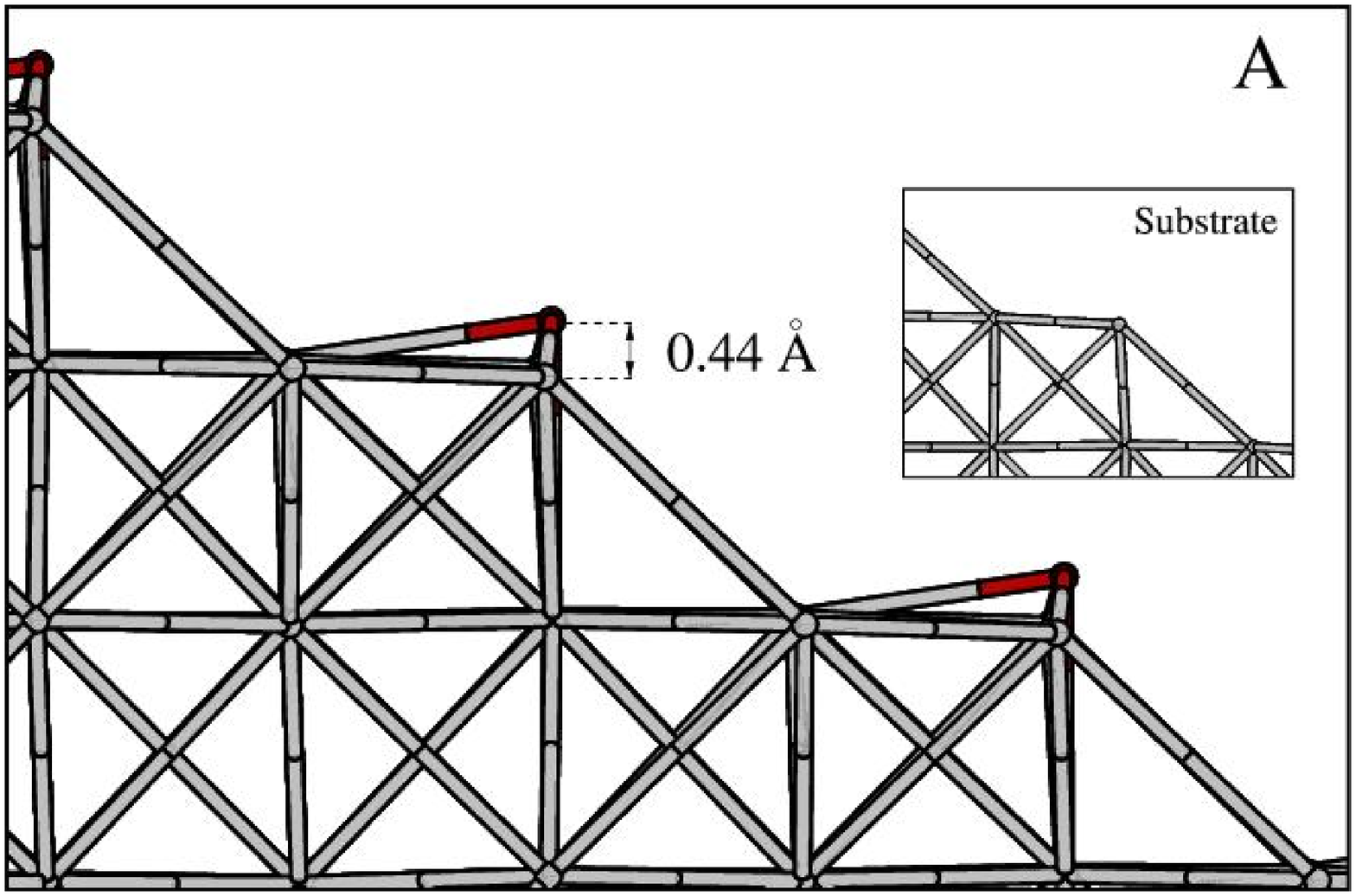}&
    \includegraphics*[width=8.0cm]{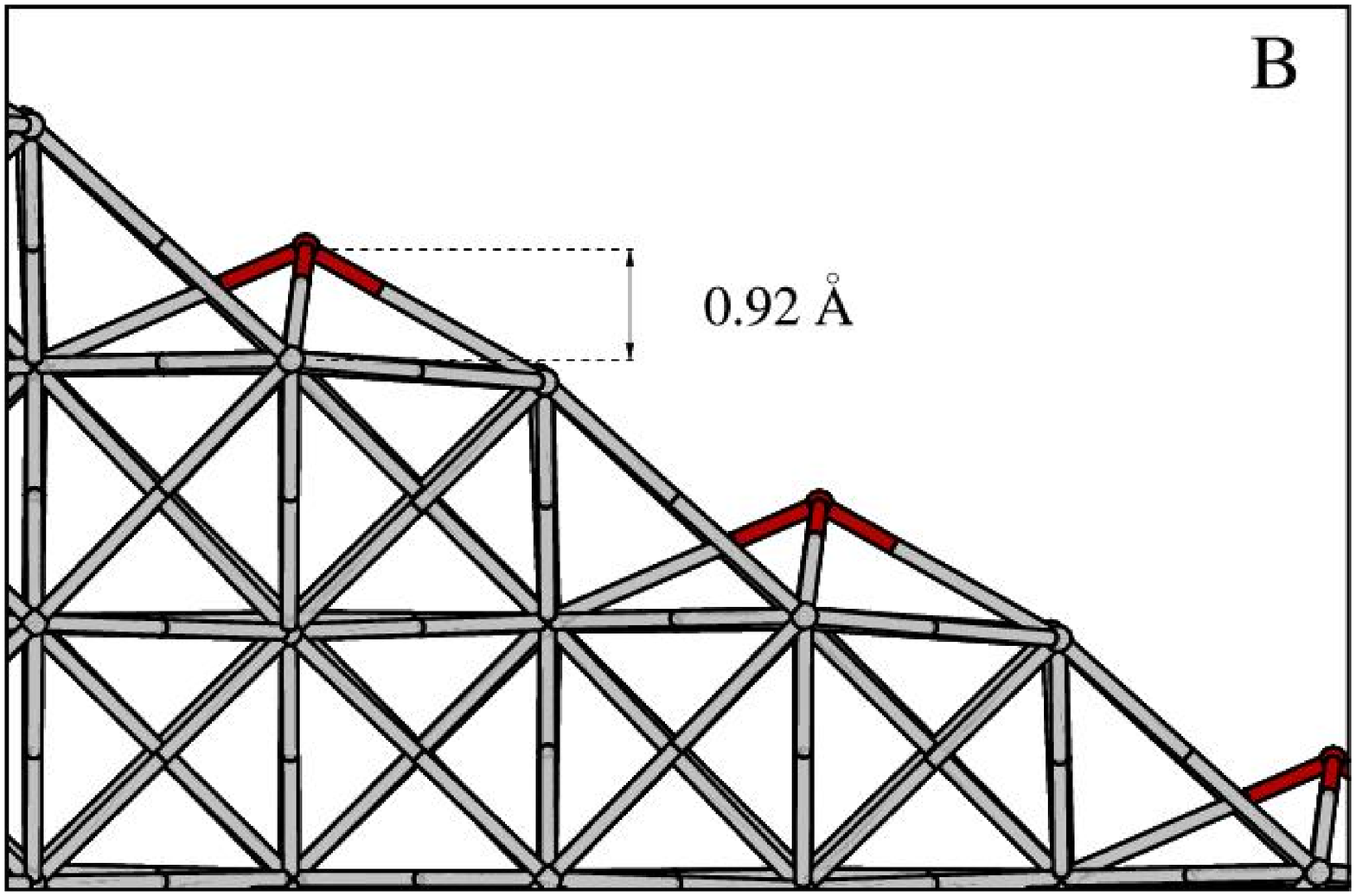}\\[0.3em]
    \includegraphics*[width=8.0cm]{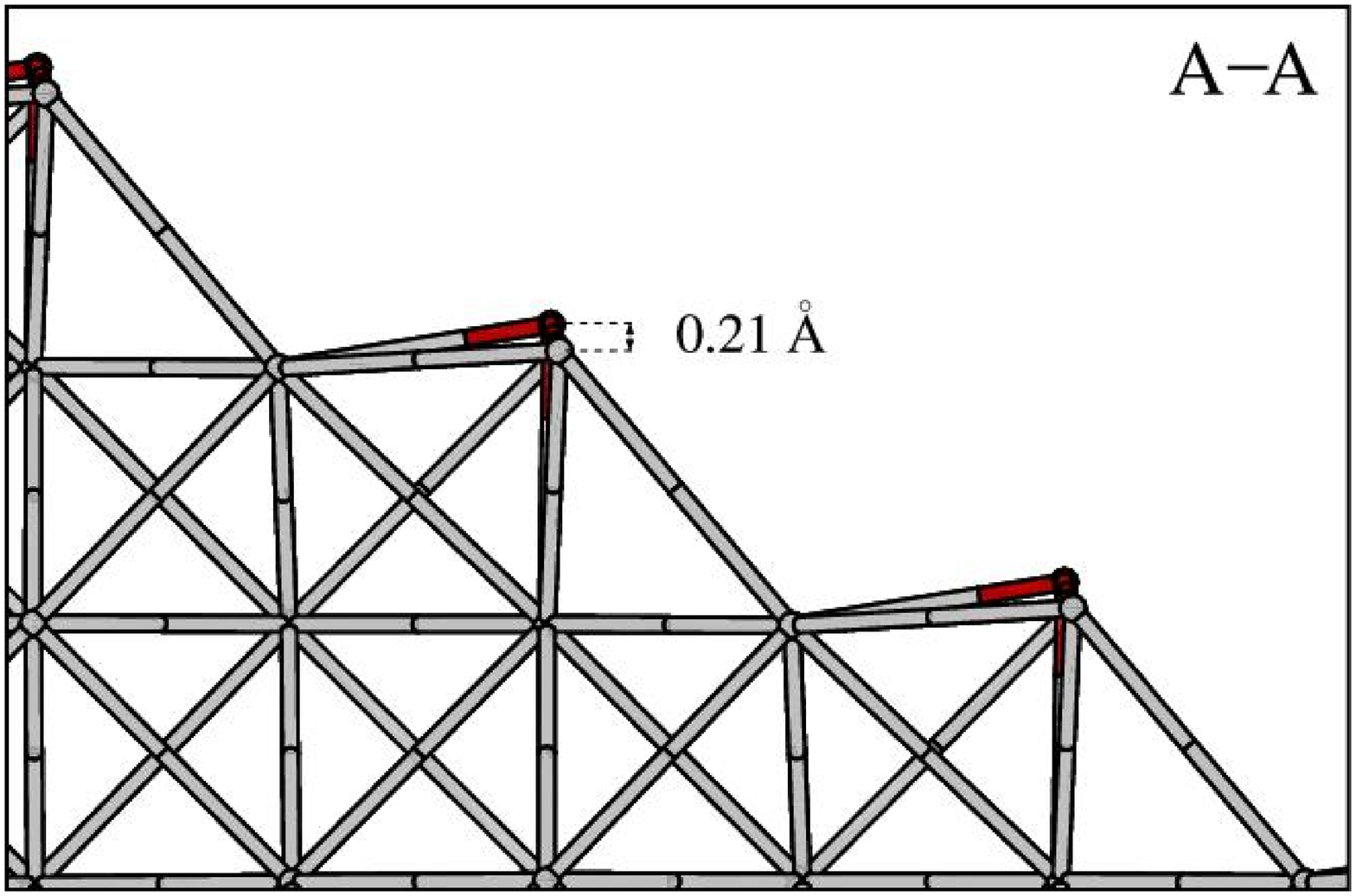}&
    \includegraphics*[width=8.0cm]{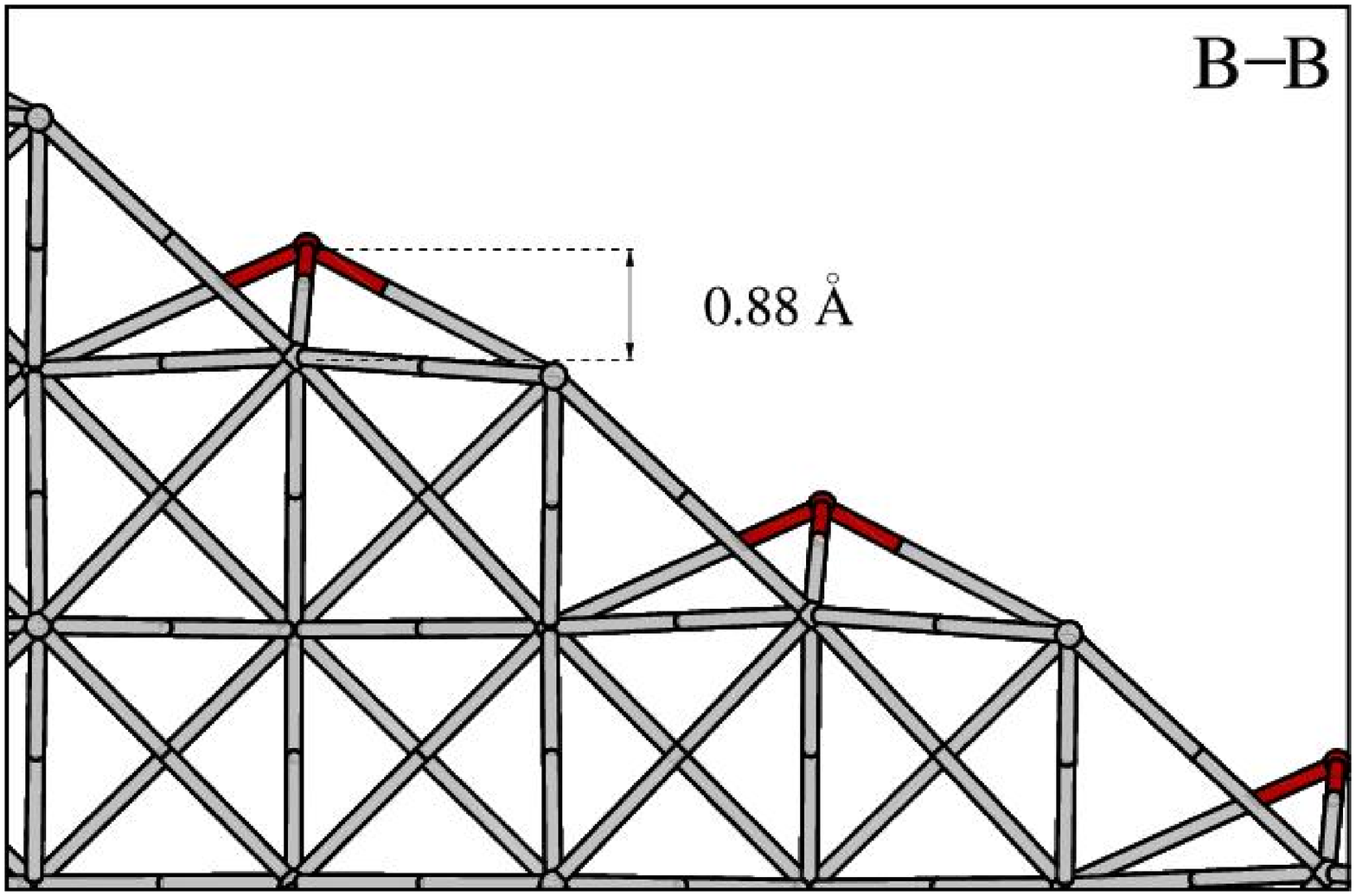}\\[0.3em]
  \end{tabular}
  \caption{(Color online) Side views of the Ag(210) substrate with O
    adatoms on site A (left panels) and on site B (right panels) in
    the (2$\times$1) and (1$\times$1) geometries. Grey (red) balls
    represent silver (oxygen) atoms. The inset in the top-left panel
    shows the side view of the optimized clean Ag(210).  }
  \label{fig:a-oag210}
\end{figure*}

In Table \ref{tab:2b} we report the optimized structural parameters
describing the oxygen adsorption on Ag(210) at $\Theta =$ 1/4 ML and
at $\Theta =$ 1/2 ML for the S$_1$--S$_1$ configurations, while
Fig. \ref{fig:a-oag210} shows the side views of the corresponding
optimized O/Ag(210) structures.  These data and plots reveal the same
trend as for Ag(410) surface. The most apparent feature is that the
height of the O adatoms above the surface is much smaller for A and
A--A sites than for the B and B--B sites. Note also the difference
between the pattern of the substrate reconstruction for A (low
coverage) and A--A (high coverage) configurations. Like on the Ag(410)
surface, the adsorption of O adatoms induces an outward relaxation of
step Ag atoms for both the A and A--A configurations. In the case of
the A configuration this relaxation is not large enough to compensate
the inward relaxation of the clean substrate (see the inset of
top-left panel of Fig.  \ref{fig:a-oag210}), therefore Ag atoms at the
step appear to be pushed somewhat inward, like in the case of the
clean Ag(210) substrate. On the contrary, this relaxation is quite
significant in the A--A configuration, resulting in a marked expansion
of the bond length between the Ag step and terrace atoms ($\Delta
d_{\rm Ag^R-Ag^{S'}} = 0.20$~\AA), and between the Ag step atom and Ag
atom located below the O adatom ($\Delta d_{\rm Ag^R-Ag^B} =
0.19$~\AA). Note that the step Ag atoms appear to be pushed
outwards. Contrary to the A--A geometry, in the A configuration the
step Ag atoms also relax laterally along the step-edge direction away
from O adatom---the distance between the two step Ag atoms that are
bound to the same O adatom being increased by 0.17 \AA.

%
%

\subsection{Electronic structure}
In order to better characterize the different behavior of oxygen on
the step in A and A--A configurations we have analyzed the
corresponding electronic structure. Here we present the analysis of
the electronic structure of O/Ag(410) only---the conclusions for
O/Ag(210) being similar.

\begin{figure}
\centerline{\includegraphics[width=9.0cm]{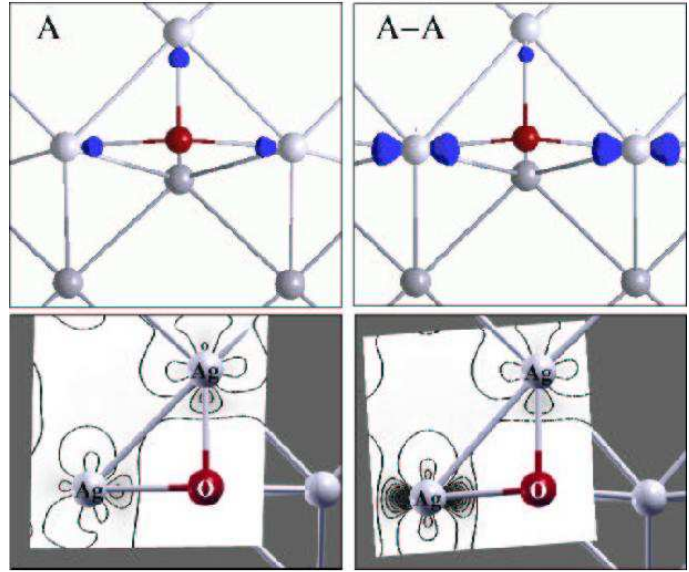}}
  \caption{(Color online) Charge difference density, i.e. the
    difference between the electron density of the O/Ag system and the
    density of the clean surface [$\Delta n (r) = n_{\rm O/Ag}(r) -
    n_{\rm Ag}(r)$], for A and A--A configurations (top views).  Top
    panels: isosurfaces at $-$0.015 e/a$_0^3$ (blue).  Bottom panels:
    charge density contours in a plane passing through O, Ag$^{\rm R}$
    and Ag$^{\rm S}$ atoms.  Contours are drawn in linear scale from
    $-0.07$ to $0.0$ e/a$_0^3$, with the increment of $0.01$
    e/a$_0^3$.  Grey (red) balls represent silver (oxygen) atoms.  }
  \label{fig:charge}
\end{figure}

The interaction between oxygen and silver removes electronic charge
from the silver atoms neighboring the O adatom.  The electron deficit
regions around the silver atoms are shown in Fig. \ref{fig:charge}
(blue isosurfaces).  It is interesting to observe that while at low
coverage the deficit regions around the three Ag atoms are very
similar, thus indicating a similar donation of charge, in the A--A
configuration the flow of charge to oxygen adatoms is mainly due to
silver atoms on the step. This suggests that the O--Ag--O chains that
form in the (1$\times$1) geometry have the structure of
electrostatically stable $\cdots +-+- \cdots$ strings.
  
\begin{figure}
\centerline{\includegraphics*[width=7.0cm]{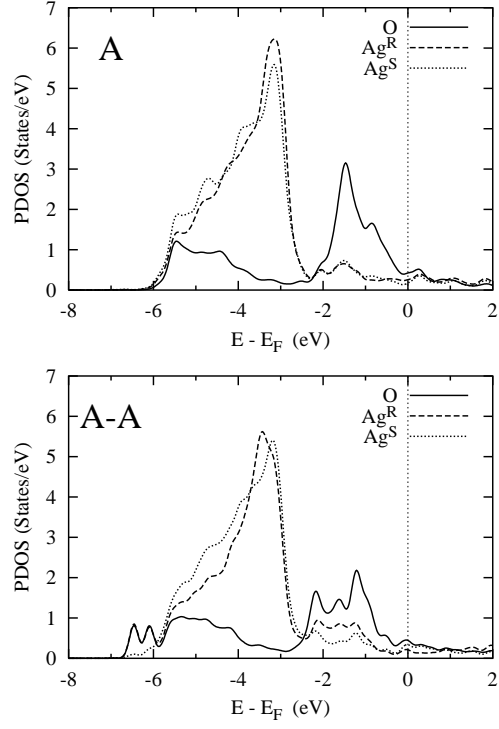}}
  \caption{Density of states projected onto oxygen and silver atoms
    for A and A--A configurations. Ag$^{\rm R}$ and Ag$^{\rm S}$ as in
    Table \ref{tab:2}.}
  \label{fig:dos}
\end{figure}

\begin{figure}
  \centerline{\includegraphics*[width=8.0cm]{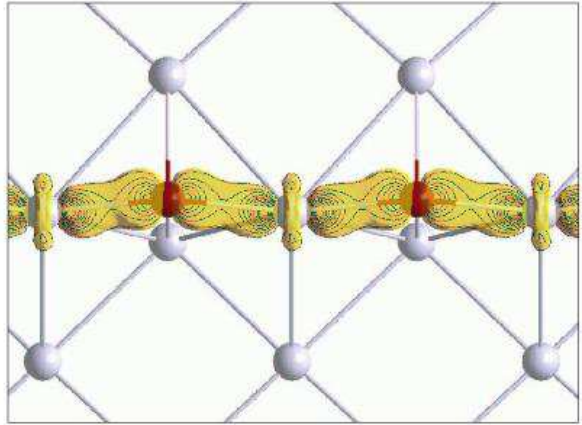}}
  \caption{(Color online) ILDOS in the energy window ($-$7.0,
    $-$6.0)eV for the A--A configuration (top view).  The isosurface
    is at 0.015 e/a$_0^3$ (yellow) and the 10 contours are drawn in
    linear scale from 0.015 e/a$_0^3$ to 0.15 e/a$_0^3$.}
  \label{fig:ildos}
\end{figure}

The analysis of the density of states reveals that the bond between
oxygen and silver is not purely ionic. In Fig.  \ref{fig:dos} we
display the density of states projected (PDOS) onto the oxygen adatom
and onto the nearest silver atoms for both the A and A--A
geometries. The PDOS on O atoms shows that the interaction of the O
2$p$ state with the 4$d$ metal band results in the formation of two
regions of high density of states, which correspond to bonding and
antibonding states.  The bonding states are located at the bottom of
the 4$d$ Ag band while the antibonding ones are mainly below the Fermi
level. Note that the O 2$s$ level lies below the bottom of the valence
band (at about $-$17 eV with respect to the Fermi level) and it is not
shown in the figures. The main difference between the PDOS in the two
geometries is that at high coverage additional features appear 6--7 eV
below the Fermi level. The figure clearly shows that these states are
due the hybridization of oxygen 2$p$ orbitals and 4$d$ orbitals of the
silver atoms on the step (Ag$^{\rm R}$). By inspection of the
components of these orbitals we find that the mixing is mainly between
O 2$p_x$ orbital parallel to the step-edge and Ag 4$d_{x^2-y^2}$
orbital parallel to the terrace plane.  This hybridization is very
efficient because of the good alignment between O and Ag atoms on the
step. The spatial distribution of these states is illustrated in Fig.
\ref{fig:ildos}, where we display the integrated local density of
states (ILDOS)\cite{koka02} in the energy window ($-$7.0, $-$6.0) eV
comprising the two small peaks in the PDOS. The figure shows quite
clearly the bond between oxygen and silver atoms on the step.  It is
worth noting that in the other S$_1$--S$_1$ configurations there are
no states in this energy region even if we find an increase of the
density of states around $-6.0$ eV, especially in the T$_2$--T$_2$
geometry.
  
The study of the electronic structure of the added-row reconstructed
Ag(110) surface confirms that the presence of bonding states 6--7 eV
below the Fermi level is related to the formation of aligned
O--Ag--O--Ag structures.  Indeed we find similar spectral features
also in the DOS of Ag(110)p(2$\times$1)O and their spatial
distribution closely resembles that shown in
Fig. \ref{fig:ildos}. Note that the results of experimental
studies\cite{cour97,seki00} show a quite weak oxygen-induced feature
in this energy range.

\subsection{Oxygen vibrational modes}
\begin{figure}
  \centerline{\includegraphics*[width=8.0cm]{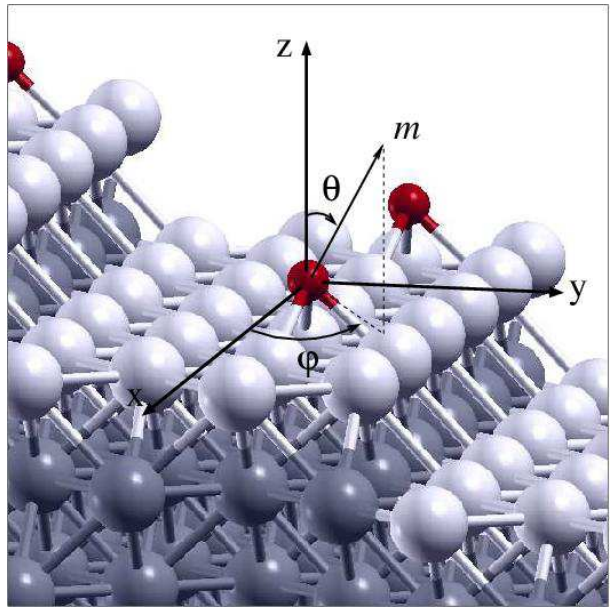}}
  \caption{(Color online) Schematic definition of the angles $\theta$
    and $\varphi$ of the displacement eigenvector $m$. Observe that
    the $z$ axis of the coordinate system is perpendicular to the
    terrace (100) plane while the $x$ axis is parallel to the step
    edge.}
  \label{fig:ref-osc}
\end{figure}

In Table \ref{tab:osc} we present the frequencies of the adsorbate
modes and the directions of the corresponding displacement
eigenvectors (the angles are defined in Fig. \ref{fig:ref-osc}) for
different O/Ag systems.  Before analyzing the vibrational modes of
oxygen on Ag(410) and Ag(210), it is useful to discuss the vibrational
properties of the O/Ag(100) system with oxygen adsorbed in hollow
sites and of the added-row reconstructed Ag(110) surface.  Data for
the Ag(100)c(2$\times$2)O structure---which had already been
calculated in Ref.~\onlinecite{loff03} in the local density
approximation (LDA)---have been recalculated within the present GGA
approach, giving quite similar results (not unusually, GGA frequencies
result to be slightly red-shifted with respect to LDA).  The $m_1$
mode perpendicular to the surface vibrates at 30 meV in agreement with
the peak at 30-32 meV observed in the HREEL spectra of the
non-reconstructed O/Ag(100) surface.\cite{loff03} The modes $m_2$ and
$m_3$ at 50 meV are two in-plane degenerate modes.  For the
Ag(110)p(2$\times$1)O structure we find a vibrational mode, $m_2$,
perpendicular to the surface at 38 meV. This value has to be compared
with the feature around 41 meV observed in HREEL
spectra.\cite{vatt94,stie94} Besides this dipole-active mode two other
adsorbate modes exist, $m_1$ at 28 meV and $m_3$ at 73 meV, in which
the adatom vibrates parallel to the surface plane.  The polarization
of the softer mode is perpendicular to the added row while the other
one is parallel to it.  It is interesting to observe that the $m_3$
mode is even harder than the in-plane modes ($m_2$ and $m_3$) of
Ag(100)c(2$\times$2)O because the oxygen atom sits closer to the
nearest silver atoms in the added row reconstructed Ag(110) surface
($\Delta z$ = 0.10 \AA) than in the Ag(100)c(2$\times$2)O structure
($\Delta z$ = 0.71 \AA)\cite{cipr02}.

\begin{figure}
  \centerline{\includegraphics*[width=8.0cm]{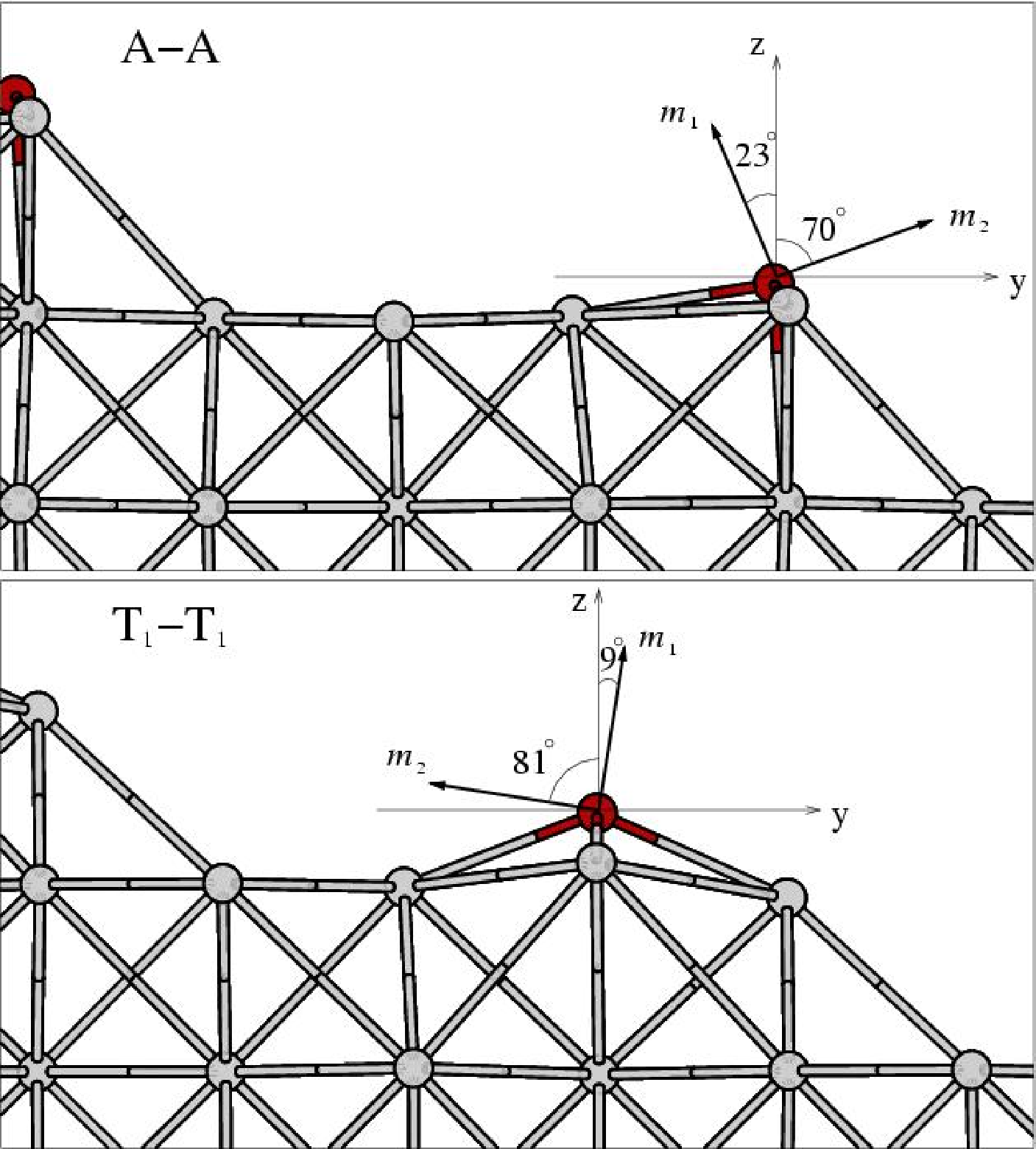}}
  \caption{(Color online) Displacement eigenvectors $m_1$ and $m_2$
    for the A-A and the T$_1$-T$_1$ configurations on Ag(410). For the
    A-A (T$_1$-T$_1$) geometry the mode $m_1$ vibrates at 24 meV (29
    meV) while the mode $m_2$ at 37 meV (32 meV).}
  \label{fig:modi}
\end{figure}

We now focus on the vibrational properties of the A-A adsorption
geometry on Ag(410) and Ag(210).  The mode at higher frequency ($m_3$
at 68 meV and 71 meV for Ag(410) and Ag(210) respectively) in which the
adatom vibrates parallel to the Ag--O--Ag chain at the step is very
similar to the $m_3$ mode of the Ag(110)p(2$\times$1)O structure.  The
softer modes are due to vibrations of the adsorbate in a plane
perpendicular to the step edge (these two modes for the Ag(410)
surface are shown in the upper panel of Fig. \ref{fig:modi}).  The
$m_1$ mode at 24 meV is nearly parallel to the (110) plane of the step
while the $m_2$ mode at 37 meV is almost perpendicular to it. These
modes resemble very closely the $m_1$ and $m_2$ modes of the
Ag(110)p(2$\times$1)O surface. The displacement pattern of these two
modes suggests that they could both be dipole active.
In particular the mode $m_2$ could be responsible of the spectral
feature around 40 meV observed in the HREEL spectra of these surfaces.

For the Ag(210) surface we have also investigated the A configuration
to understand the effect of the Ag--O--Ag chain formation at the step
on the vibrational properties. In this case both the frequency and the
displacement eigenvectors of the softer modes, $m_1$ and $m_2$, change
very little with respect to the A--A arrangement. The $m_3$ mode is
polarized parallel to the step and is significantly softer than in the
A--A structure (46 versus 71 meV) mainly because in this case the
height of the adatom with respect to the step Ag atoms is larger than
in the A--A geometry.

Since the hollow terrace sites on Ag(410) compete in energy with the
step edge sites, we have also investigated the vibrational properties
of a very stable adsorption configuration which involves terrace
sites, namely the T$_1$--T$_1$ geometry. In the lower panel of Fig.
\ref{fig:modi} we show the $m_1$ and $m_2$ modes in the plane
perpendicular to the step. The $m_1$ mode at 29 meV is due to the
vibration of O nearly perpendicular to the (100) terrace plane while
the $m_2$ mode at 32 meV is due to O vibrating almost parallel to this
plane. The harder mode, $m_3$, is at 58 meV and it corresponds to the
oxygen vibration parallel to the step edge.  Our calculations show
clearly that the $m_1$ mode is very similar to the dipole active $m_1$
mode of the c(2$\times$2) O adsorption geometry on Ag(100). It seems
thus reasonable to expect that the spectral feature around 32 meV
observed in HREEL spectra on Ag(410) could be due to this stretching
mode.

Finally we have calculated the vibrational modes for the B--B
adsorption geometry on Ag(210). An adatom at site B is significantly
less stable than at the step edge, but it is interesting to
investigate its vibrational properties since this is the only terrace
site available on Ag(210) and the HREEL spectra of this surface show,
in addition to the 40 meV peak, other spectral features around 30 meV
and 56 meV.  As it is shown in Table \ref{tab:osc}, the higher
frequency mode, $m_3$ at 37 meV, is parallel to the step edge, while
the other two modes are in the plane perpendicular to it. In
particular the $m_1$ mode at 27 meV is nearly perpendicular the
(110) step plane, while the other, $m_2$ at 34 meV, is almost parallel
to it.

Two important results emerge from this analysis.
The $m_1$ mode in the T$_1$-T$_1$ configuration of O/Ag(410) at 29 meV
lies very close to the $m_1$ mode of O/Ag(100) (30 meV), in
correspondence with the spectral feature around 32 meV observed in
HREEL spectra on Ag(410).\cite{savi01,savi02,vatt03} The $m_2$ modes
of the A-A configuration in O/Ag(210) and O/Ag(410) (both at 37 meV)
lie close to the $m_2$ frequency in the added-row O/Ag(110) surface
[Ag(110)p(2$\times$1)O] at 38 meV, in correspondence with the spectral
feature around 40 meV observed in the HREEL spectra of O/Ag(410) and
O/Ag(210) surfaces.\cite{savi01,savi02,vatt03} We believe that this
correspondence between infrared-active vibrational modes, calculated
for specific configurations, and the features found in the
experimental HREEL spectra supports the proposed assignment of those
spectral features to atomic species adsorbed on terraces and at step
edges. Problems still remain in the interpretation of the HREEL
spectra of O/Ag(210).
On one hand, the feature experimentally found at 32 meV could be due
to the vibration of oxygen occupying the only available terrace
sites (site B), but this assignment remains uncertain since the
adsorption on these sites is energetically unfavored.
On the other hand, no vibration is predicted near 56 meV, where a
feature in the experimental spectrum is also found.  This feature
could be the signature of other O species, possibly adsorbed
subsurface, which we have not investigated in the present work. For
sure, these issues call for further theoretical and, possibly,
experimental work.

\section{Conclusions}
In this paper we have investigated the adsorption of oxygen on Ag(410)
and Ag(210) surfaces. Our results show that the step edge decoration
is particularly stable on both surfaces. While on Ag(410) adsorption
on terrace and at the step edge compete in energy and no clear
preference is predictable, on Ag(210) the step decoration is
significantly favored with respect to other adsorption configurations.
The formation of this stable O--Ag--O rows at the step edges strongly
affects both the structural and the electronic properties of the
surface, and leaves a clear fingerprint in the HREEL spectrum of these
systems at 40 meV.

The situation is less clear for the HREEL peak at 32 meV. On Ag(410)
it is reasonable to assign this spectral feature to adatoms occupying
terrace sites.  On Ag(210) instead this O--Ag stretch mode could be
due to O atoms adsorbed on terraces at the step foot, which however
are predicted to be metastable adsorption sites. Among the adsorption
configurations that we have investigated, no candidate was found to
support the vibrational mode experimentally found at 56 meV. This
suggests that other O species, possibly adsorbed subsurface, would be
responsible for this spectral feature.

\section*{Acknowledgments}
This work has been supported by INFM ({\it Iniziativa trasversale
calcolo parallelo, Sezioni F e G}, and {\it PAIS Chemde}) and by the
Italian MIUR through PRIN. All numerical calculations were performed
at the CINECA national supercomputing center in Bologna (Italy).

\newpage
%
%

%
%
\begin{table}
  \begin{center} 
    \caption{Chemisorption energies, $E_{\rm chem}$, and O--O distances,
      $d_{{\rm O}-{\rm O}}$, for various chemisorption sites on Ag(410)
      surface.  Values in parentheses are obtained keeping the substrate
      fixed.}\label{tab:1}
    \begin{ruledtabular}
      \begin{tabular}{l l c l}
        $\Theta$ (ML)  & Configuration & $d_{\rm O-O}$ (\AA) &
        $E_{\rm chem}$ (eV) \\
        \hline
            & A            & 8.32 &  $-$0.75  ($-$0.66) \\
        1/8 & T$_1$        & 8.32 &  $-$0.80            \\
            & T$_2$        & 8.32 &  $-$0.83  ($-$0.76) \\
            & B            & 8.32 &  $-$0.60            \\
        \hline
            & A--A         & 4.16 &  $-$0.86  ($-$0.70) \\
            & T$_1$--T$_1$ & 4.16 &  $-$0.78            \\
            & T$_2$--T$_2$ & 4.16 &  $-$0.79  ($-$0.69) \\
            & B--B         & 4.16 &  $-$0.54            \\
            &              &      &                     \\
        1/4 & A--T$_2$     & 4.23 &  $-$0.80            \\
            & T$_1$--B     & 4.22 &  $-$0.63            \\
            &              &      &                     \\
            & A--T$_1$     & 3.26 &  $-$0.59            \\
            & T$_1$--T$_2$ & 3.22 &  $-$0.66            \\
            & T$_2$--B     & 3.23 &  $-$0.54            \\
            & A--B         & 3.30 &  $-$0.56            \\
      \end{tabular} 
    \end{ruledtabular}
  \end{center}
\end{table}

%
%
\begin{table}
  \begin{center} 
    \caption{Chemisorption energies and O--O distances for various 
      chemisorption sites on Ag(210) surface. The labels have the same
      meaning as in Tab. \ref{tab:1}.  }
    \label{tab:1b}
    \begin{ruledtabular}
      \begin{tabular}{l l c c}
        $\Theta$ (ML) & Configuration & $d_{\rm O-O}$ (\AA) &  $E_{\rm
          chem}$ (eV) \\ 
        \hline
        1/4  & A     & 5.10 &  $-$0.68  \\
             & B     & 5.10 &  $-$0.49  \\
        \hline              
             & A--A  & 4.16 &  $-$0.80  \\
        1/2  & B--B  & 4.16 &  $-$0.42  \\
             &       &      &           \\
             & A--B  & 3.62 &  $-$0.56  \\
      \end{tabular} 
    \end{ruledtabular}
  \end{center}
\end{table}

%
%

\begin{table*}[t]
  \begin{center}
    \caption{
      Some key quantities describing the O chemisorption sites on Ag(410) 
      in various geometries. $d_{\rm O-Ag^{\rm R}}$ is bond length with 
      the nearest silver atom, Ag$^{\rm R}$ see Fig. \ref{fig:legend}, 
      lying in the row of oxygen adatoms. $d_{\rm O-Ag^S}$ and $d_{\rm 
        O-Ag^{S'}}$ indicate the bond length with the other two oxygen
      nearest surface silver atoms, Ag$^{\rm S}$ and Ag$^{\rm S'}$, 
      while and $d_{\rm O-Ag^B}$ is the bond length with the silver
      atom just below oxygen, Ag$^{\rm B}$. $\Delta z$ is the height
      of oxygen with respect to Ag$^{\rm R}$ atoms and $\alpha$ is the
      bond angle between the bonds with the nearest Ag$^{\rm R}$ atoms
      ($\alpha = 180^\circ$ indicates perfect alignment).  
    }
    \label{tab:2}
    \begin{ruledtabular}
      \begin{tabular}{l l  c c c c c c }
        $\Theta$ (ML) & Configuration     & $d_{\rm O-Ag^{\rm R}}$ (\AA) & 
        $d_{\rm O-Ag^S}$ (\AA)
        & $d_{\rm O-Ag^{S'}}$ (\AA)& $d_{\rm O-Ag^B}$ (\AA)&  $\Delta z$ (\AA)&
        $\alpha$ ($^\circ$) \\
        \hline
            & A            & 2.21   &      & 2.21 & 2.43 & 0.46  & 156  \\
        1/8 & T$_1$        & 2.27   & 2.26 & 2.29 & 2.97 & 0.80  & 138  \\
            & T$_2$        & 2.28   & 2.25 & 2.25 & 2.99 & 0.82  & 138  \\
            & B            & 2.30   & 2.24 & 2.37 & 3.08 & 0.88  & 134  \\
        \hline
            & A--A         & 2.10   &      & 2.24 & 2.40 & 0.24  & 164  \\
        1/4 & T$_1$--T$_1$ & 2.15   & 2.31 & 2.30 & 2.91 & 0.55  & 150  \\
            & T$_2$--T$_2$ & 2.17   & 2.28 & 2.29 & 2.92 & 0.60  & 148  \\
            & B--B         & 2.20   & 2.27 & 2.37 & 2.99 & 0.71  & 142  \\
        \hline
            & Ag(110)p(2$\times$1)O & 2.08 & &   &       & 0.10  & 175  \\
      \end{tabular}
    \end{ruledtabular}
  \end{center}
\end{table*}

%
%
\begin{table*}
  \begin{ruledtabular}
    \caption{Some key quantities of O/Ag(410) structures 
      describing the bond lengths between
      Ag$^{\rm R}$ and the nearest silver atom for various configurations
      (Ag$^{\rm R}$, Ag$^{\rm S}$ and Ag$^{\rm B}$ as in Tab. \ref{tab:2}). 
      The distances $d_{\rm Ag^X-Ag^Y}$ always refer to the distance 
      between the two given silver atoms that are bonded to the
      same O adatom.
      The values in parentheses refer to the clean Ag(410) surface.
    }
    \label{tab:3}
    \begin{tabular}{l l  l l l l  }
      $\Theta$ (ML) & Configuration & $d_{\rm Ag^{\rm R}-Ag^{\rm R}}$ (\AA) &
      $d_{\rm Ag^{\rm R}-Ag^S}$ (\AA) & $d_{\rm Ag^{\rm R}-Ag^{S'}}$ (\AA)&
      $d_{\rm Ag^{\rm R}-Ag^B}$ (\AA) \\
      \hline
          & A            & ~~4.32 (4.16) &               & ~~3.08 (2.90) & 
                                                           ~~2.93 (2.84)    \\
      1/8 & T$_1$        & ~~4.26        & ~~2.97 (2.91) & ~~2.98 (2.90) & 
                                                           ~~3.07 (2.91)    \\
          & T$_2$        & ~~4.26        & ~~2.98 (2.92) & ~~3.00 (2.92) & 
                                                           ~~3.04 (2.92)    \\
          & B            & ~~4.24        & ~~2.99 (2.92) & ~~2.98 (2.91) & 
                                                           ~~3.06 (2.96)    \\
      \hline
          & A--A         & ~~4.16        &               & ~~3.15        & 
                                                           ~~3.00           \\
      1/4 & T$_1$--T$_1$ & ~~4.16        & ~~3.00        & ~~2.99        & 
                                                           ~~3.15           \\
          & T$_2$--T$_2$ & ~~4.16        & ~~2.99        & ~~2.99        & 
                                                           ~~3.12           \\
          & B--B         & ~~4.16        & ~~3.02        & ~~2.98        & 
                                                           ~~3.08
    \end{tabular}
  \end{ruledtabular}
\end{table*}

%
%
\begin{table*}
  \caption{
    Some key quantities describing the O chemisorption sites on
    Ag(210) in various geometries. The labels have the same meaning as
    in Tab. \ref{tab:2}.
  }
  \label{tab:2b}
  \begin{ruledtabular}
    \begin{tabular}{l l c c c c c c }
      $\Theta$ (ML)            & Configuration          & 
      $d_{\rm O-Ag^R}$ (\AA)   & $d_{\rm O-Ag^S}$ (\AA) & 
      $d_{\rm O-Ag^{S'}}$ (\AA)& $d_{\rm O-Ag^B}$ (\AA) &  
      $\Delta z$ (\AA)         & $\alpha$ ($^\circ$) \\
      \hline
      1/4 & A      & 2.21  &      & 2.21 & 2.42  & 0.44  & 158  \\
          & B      & 2.31  & 2.23 & 2.40 & 3.11  & 0.92  & 132  \\
      \hline
      1/2 & A--A   & 2.09  &      & 2.24 & 2.42  & 0.21  & 168  \\
          & B--B   & 2.26  & 2.25 & 2.41 & 3.09  & 0.88  & 134  \\
    \end{tabular} 
  \end{ruledtabular}
\end{table*}

%
%
\begin{table*}
\caption{
  Some key quantities of O/Ag(210) structures describing the bond
  lengths between Ag$^{\rm R}$ and the nearest silver atom for various
  configurations (Ag$^{\rm R}$, Ag$^{\rm S}$, Ag$^{\rm S'}$ and
  Ag$^{\rm B}$ as in Tab. \ref{tab:2}). The distances $d_{\rm
    Ag^X-Ag^Y}$ always refer to the distance between the two given
  silver atoms that are bonded to the same O adatom. The values in
  parentheses refer to the clean Ag(210) surface. 
}
\label{tab:3b}
\begin{ruledtabular}
\begin{tabular}{l l l l l l }
$\Theta$ (ML)                & Configuration              & 
$d_{\rm Ag^R-Ag^R}$ (\AA)    & $d_{\rm Ag^R-Ag^S}$ (\AA)  & 
$d_{\rm Ag^R-Ag^{S'}}$ (\AA) & $d_{\rm Ag^R-Ag^B}$ (\AA)  \\
\hline
1/4 & A    & ~~4.33 (4.16) &               & ~~3.04 (2.89) & ~~2.95 (2.85) \\
    & B    & ~~4.22        & ~~2.96 (2.89) & ~~2.97 (2.91) & ~~3.04 (2.95) \\
\hline
1/2 & A--A & ~~4.16        &               & ~~3.09        & ~~3.04        \\
    & B--B & ~~4.16        & ~~2.95        & ~~2.97        & ~~3.03        \\
\end{tabular} 
\end{ruledtabular}
\end{table*}

%
%
\begin{table*}
  \caption{
    Vibrational frequencies and displacement eigenvectors for
    different O/Ag structures. The angles which define the directions
    of the eigenvectors are shown in Fig. \ref{fig:ref-osc}. For both
    the Ag(110)p(2$\times$1)O and O/Ag(100)-c(2$\times$2) systems
    $\theta$ is the angle with respect to the axis perpendicular to
    the surface. For the Ag(110)p(2$\times$1)O $\varphi$ is the angle
    in the surface plane with respect to the added row. The values in
    parentheses are the LDA results.\cite{loff03} 
  }
  \label{tab:osc}
  \begin{ruledtabular}
    \begin{tabular}{l l l c c c}
      Surface & Configuration & Mode & $\nu$ (meV) & $\theta$ ($^\circ$)&
      $\varphi$  ($^\circ$)\\ 
      \hline
      &               & $m_1$ & 30 (35)  & 0  &  -  \\
      Ag(100) & c(2$\times$2) & $m_2$ & 50 (55)  & 90 &  -  \\
      &               & $m_3$ & 50 (55)  & 90 &  -  \\
      \hline
      &               & $m_1$ & 28  & 90 & 90  \\
      Ag(110) & added row     & $m_2$ & 38  & 0  &  -  \\
      &               & $m_3$ & 73  & 90 &  0  \\
      \hline
      &               & $m_1$ & 24  & 23 & 270 \\
      & A--A           & $m_2$ & 37  & 70 &  90 \\
      &               & $m_3$ & 68  & 90 &  0  \\
      &               &       &     &    &     \\
      Ag(410)         & & $m_1$ & 29  &  9 &  90 \\
      & T$_1$--T$_1$   & $m_2$ & 32  & 81 & 270 \\
      &               & $m_3$ & 58  & 90 &  0  \\
      \hline
      &               & $m_1$ & 24  & 23 & 270 \\
      & A--A           & $m_2$ & 37  & 69 &  90 \\
      &               & $m_3$ & 71  & 90 &  0  \\
      &               &       &     &    &     \\
      &               & $m_1$ & 25  & 30 & 270 \\
      Ag(210) & A             & $m_2$ & 40  & 64 &  90 \\
      &               & $m_3$ & 46  & 90 &  0  \\
      &               &       &     &    &     \\
      &               & $m_1$ & 27  & 39 &  90 \\
      & B--B           & $m_2$ & 34  & 49 & 270 \\
      &               & $m_3$ & 37  & 90 &  0  \\
    \end{tabular}
  \end{ruledtabular}
\end{table*}

\newpage
%
%

%
%

%
%

%
%

%
%

%
%

%
%

%
%

%
%

%
%

%
%

%
%

%
%


\begin{thebibliography}{99}
\bibitem{kmesh} 
    The BZ integrations have been performed using the uniformly
    spaced and shifted $k$-meshes. Such a mesh is generated by the
    following recipe: (i) define the sequence of numbers $u_i =
    \frac{r-1}{q} + \frac{1}{2q}, r=1 \ldots q$; and (ii) generate the
    following $q^3$ distinct $\mathbf{k}$ points: $\mathbf{k}_{ijk} =
    u_i\mathbf{a}^* + u_j\mathbf{b}^* + u_k\mathbf{c}^*$, where
    $\mathbf{a}^*$, $\mathbf{b}^*$, and $\mathbf{c}^*$ are reciprocal
    lattice vectors. It can be shown, taking into account the
    tranlational symmetry, that in case of even mesh (i.e. even set of
    $\{u_l\}_{l=i,j,k}$ values) such mesh coincides with the
    Monkhorst-Pack mesh. \cite{monk76}
\bibitem{cover}  
  The coverage is defined as the number of oxygen atoms
  divided by the number of Ag surface atoms in the unit-cell.
\bibitem{labels-note} 
  Although the Ag($n$10) substrate possesses the
  mirror-plane symmetry, which is parallel to [$1\overline{n}0$] crystal
  axis and perpendicular to [001] axis (i.e. the plane passes through
  the Ag$^{\rm S}$ and Ag$^{\rm S'}$ atoms in Fig. \ref{fig:legend}), we
  have not imposed this symmetry in our calculations. Therefore the two
  R atoms in Fig. \ref{fig:legend} are located in slightly asymmetric
  positions, and in Tabs. III--VI the distances labeled as $d_{\rm
    O-Ag^{R}}$, $d_{\rm Ag^{R}-Ag^{S}}$, $d_{\rm Ag^{R}-Ag^{S'}}$, and
  $d_{\rm Ag^{R}-Ag^B}$ are averaged distances.
\end{thebibliography}
\end{document}